\documentclass[manuscript,screen,acmsmall]{acmart}

\AtBeginDocument{%
  \providecommand\BibTeX{{%
    \normalfont B\kern-0.5em{\scshape i\kern-0.25em b}\kern-0.8em\TeX}}}

\setcopyright{acmlicensed}
\copyrightyear{2024}
\acmYear{2024}
\acmDOI{XXXXXXX.XXXXXXX}

\acmConference[CSCW'24]{conference on Human Factors in Computing Systems}{Nov 9--13, 2024}{San José, Costa Rica}

\acmISBN{978-1-4503-XXXX-X/18/06}

\begin{document}

\title{``If it has an exclamation point, I step away from it, I need facts, not excited feelings\textit{'}': \\ Technologically Mediated Parental COVID Uncertainty}


\author{Karen Joy}
\affiliation{%
  \institution{Rutgers University}
  \city{New Brunswick,NJ}
  \country{US}}
\email{tawfiq.ammari@rutgers.edu}

\author{Michelle Liang}
\affiliation{%
  \institution{Rutgers University}
  \city{New Brunswick,NJ}
  \country{US}}
\email{tawfiq.ammari@rutgers.edu}

\author{Tawfiq Ammari}
\affiliation{%
  \institution{Rutgers University}
  \city{New Brunswick,NJ}
  \country{US}}
\email{tawfiq.ammari@rutgers.edu}

\renewcommand{\shortauthors}{Joy et al.}

\begin{abstract}
As a novel virus, COVID introduced considerable uncertainty into the daily lives of people all over the globe since late 2019. Relying on twenty-three semi-structured interviews with parents whose children contracted COVID, we analyzed how the use of social media moderated parental uncertainty about the symptoms, prognosis, long-term potential health ramifications of infection, vaccination, and other issues. We framed our findings using Mishel's Uncertainty in Illness theory. We propose new components to the theory that account for technological mediation in uncertainty. We also propose design recommendations to help parents cope with health uncertainty using social media. 
\end{abstract}



\received{10 January 2024}
\received[revised]{10 October 2024}
\received[accepted]{10 December 2024}

\maketitle

\section{Introduction}

Uncertainty in the context of illnesses is a multifaceted experience that individuals and their families often encounter in healthcare \cite{mast1995adult,mishel1999uncertainty}. It involves the lack of predictability and clarity across various dimensions of an illness. 

This uncertainty emerges from factors such as the diagnosis, prognosis, severity, and progression of the condition, presenting emotional challenges \cite{moos1984crisis}. The concept of uncertainty is defined as the inability to determine the meaning of illness-related events in situations where decisive values cannot be assigned, and outcomes are unpredictable due to insufficient cues \cite{mishel1981measurement, mishel1990reconceptualization, mishel1999uncertainty, mishel1988uncertainty}.

For parents especially, when it comes to their child's health, uncertainty leads to heightened anxiety levels, impacting various aspects of life \cite{10.1145/3479546, nelson2010parenting}. The highest levels of uncertainty are found in sudden and novel illness events or in chronic conditions where the disease course is unpredictable \cite{wright2009illness,pehora2015parents}. COVID was one such novel and unprecedented “illness”. A prevalent theory regarding uncertainty is the Uncertainty in Illness theory (UIT) \cite{mishel1981measurement,mishel1988uncertainty,mishel1999uncertainty, mishel1990reconceptualization}, initially proposed for acute illnesses and later re-conceptualized to encompass chronic illnesses, highlighting how uncertainty extends beyond medical aspects to impact personal and social dimensions of daily life \cite{zhang2017uncertainty, wright2009illness}. 

Lack of comprehensive information regarding COVID, especially COVID in children, contributed to uncertainty among parents, new parents, and parents whose children had medical complexities where they brought about changes in daily life, adding another layer of uncertainty. For this reason, parents took to online sources to find adequate support and navigate challenges effectively \cite{richards_et_al_22}. 

Given the importance of online communities for parents as they make sense of an uncertain situation, we ask: 

\begin{quote}
   \textbf{ RQ: How is parental COVID uncertainty mediated through social media?}
\end{quote}

Analyzing twenty-three semi-structured interviews with parents using a socio-technical lens, we examined how parents used social media to find information and social support as they tried to make sense of and manage uncertainties they felt about COVID. This included searching for a sense of the prognosis of the disease when contracted by their children, especially if they had special needs. Parents were also unsure about the efficacy of COVID vaccines and how or if they might have any negative effects on their children. 

While we found that various social media sites were sources of evidence-based and experiential information, online communities were not always a safe space or a site of support for their uncertainties. For example, some of our parents felt they were being led “down a rabbit hole.” Others thought that platform-level moderation choices limited or improved their access to information. Some parents felt that they were engaged in challenging conversations with family members because of  online information influences. Finally, the deluge of contradictory information increased parents' uncertainty instead of reducing it. 

In this study, we focused on how the socio-technical nature of online communities factored into how parents managed their uncertainty. Specifically, we studied the following properties of social media sites used by parents: (1) algorithmic recommendations; (2) moderation governance; and (3) the nature of online communities (e.g., hyper-localized communities, etc). Based on this analysis, we proposed two new components to Mishel’s Uncertainty in Illness theory. Our first component is a socio-technical antecedent factor to be added to the existing biological, psychological, and social factors. Based on the appraisal of information through social media, we also propose a new negative adaptation component. While the original theory did not account for a negative adaptation, we found that when parents are “burned out” of their online interaction, they infact do not manage their COVID uncertainty at all. 

\section{Related Work}
\label{sec:related_work}
Following the review of the Uncertainty in Illness theory (UIT), we review parents' social media use. Finally, we analyze how COVID has affected people's use of online communities. 

\subsection{Uncertainty in Illness Theory (UIT)}
\label{sec:UIT}

Uncertainty is a fundamental aspect of the human journey. Uncertainty has been defined in several ways: a complex cognitive stressor, a sense of loss of control, and a perceptual state or attitude of doubt or not knowing that changes over time \cite{mast1995adult, penrod2001refinement, wiener1993coping}.
It becomes particularly prominent when individuals face a life-threatening or chronic illness \cite{wright2009illness, mast1995adult, mishel1999uncertainty}. The theory of uncertainty in illness (UIT) was first proposed by Mishel as a way to understand adjustment to acute illness \cite{mishel1981measurement,mishel1990reconceptualization,mishel1999uncertainty, mishel1988uncertainty}. The theory was then re-conceptualized specifically looking at chronic illnesses and novel diseases in which uncertainty likely evolves progressively \cite{wright2009illness, mishel1990reconceptualization,clayton2018theories,bailey2017uncertainty}. UIT has its strongest support among subjects who are experiencing the acute phase of illness or are in a downward illness trajectory \cite{mishel1981measurement,mishel1990reconceptualization, mishel1999uncertainty, mishel1988uncertainty}.  UIT, in its original theory, as well as its adaptations by various studies, is illustrated over 3 main segments: (a) Antecedents, (b) Appraisal, and (c) Adaptation. The theory is presented in Figure \ref{fig:theory}. Please note that we are modifying the theory in this study. Our modifications are highlighted with dotted outlines. They will be described in detail in \S\ref{sec:discussion_uit}. 

\subsubsection{Antecedents}
\label{sec:antecedents}
Antecedents are factors that contribute to the experience of uncertainty – factors such as the complexity of the illness, lack of information, and the unpredictability of the disease course \cite {mishel1981measurement,mishel1990reconceptualization,mishel1999uncertainty, mishel1988uncertainty}. Studies claim that for uncertainty to predict the outcomes of individuals coping with illness, it is imperative to understand the antecedents of the same \cite{wallace2005finding, kang2011relationships}. 

Antecedents have three components: (1) Biological factors; (2) Psychological factors; and (3) Social factors, as proposed in the readaptation of Mishel’s theory 1988 \cite{mishel1988uncertainty} by Wright et al., 2009 \cite{wright2009illness}.  

Biological factors include (a) illness severity, (b) event familiarity, and (c) symptom pattern. An illustration of this is studied for rheumatoid arthritis (RA) patients where lack of consistent symptom patterns was the greatest predictor of uncertainty \cite{braden1987antecedents}. In a study conducted by Kang et al., researchers operationalized and tested antecedents of social support and education as structure providers, in addition to the symptom pattern, in patients with atrial fibrillation \cite{kang2005effects, kang2005relationships,kang2011relationships}. The study revealed that symptom severity emerged as the most potent predictor of uncertainty, while education and social support, as structure provider variables \cite{mishel1981measurement}, mitigated uncertainty \cite{kang2005effects,kang2005relationships,kang2011relationships,nelson2010parenting}.

Psychological factors focus on (a) learned helplessness, (b) a sense of mastery, and (c) locus of control. All three factors refer to the personal sense of self during the ongoing illness trajectory. Living in a cycle of disempowerment can lead to learned helplessness \cite{seligman1972learned,abramson1978learned,ammarimoderation}, where an individual’s future outcomes continue to suffer because they no longer believe they can control or change their future and thus, stop trying. On the other hand, if one has a sense of mastery and sees themselves as the locus of control, they think they have more control over their health and other forces that affect their life \cite{seligman1972learned,kennedy1998assessment,ammarimoderation}. Both are important components of psychological health and well-being across the lifespan \cite{mirowsky2007life,pearlin1981stress,shanahan2004developmental,thoits1995stress}.

While psychological and biological factors can be more personal in nature, the social support component focuses on the outward social context of the patient. The presence and quality of support from family, friends, and one’s broader social network can impact how individuals manage uncertainty \cite{kennedy1998assessment,albrecht2003social}.  Adequate social support can provide information, emotional assistance, and a sense of security, reducing uncertainty \cite{mishel1981measurement,mishel1990reconceptualization,mishel1999uncertainty,mishel1988uncertainty}.
Among Iranian people living with human immunodeficiency virus (HIV) and acquired immune deficiency syndrome (AIDS), those who had higher perceived social support suffered less uncertainty \cite{sajjadi2015correlation}. Widely, people with stronger social relationships have been found to have an increased likelihood of survival compared to those with weaker social relationships, with social relationships being comparable to smoking and alcohol consumption when measuring the influence on the risk of death \cite{holt2010social,cohen2004social}.

Additional social factors of importance include one’s level of education, one’s level of health literacy, and credible authority \cite{wright2009illness,mishel1981measurement,mishel1990reconceptualization,mishel1999uncertainty,mishel1988uncertainty}. A person’s level of education and health literacy can influence how they interpret and navigate uncertainty. Higher education levels can empower individuals to comprehend medical information better, reducing uncertainty \cite{wright2009illness,mishel1981measurement,mishel1990reconceptualization,mishel1999uncertainty, zhang2017uncertainty, mishel1988uncertainty}. Credible authority is another social factor that describes how access to information from credible sources, such as healthcare professionals and reputable authorities, can contribute to a person's understanding of their illness and alleviate uncertainty \cite{pernice2021we,roth2021mediated,kington2021identifying}. Trustworthy guidance from credible figures can shape the individual's appraisal of uncertainty \cite{heathcote2021symptom}. Moreover, UIT has expanded through research studies focusing on credible authority and social support, with investigators in nursing and health communication utilizing the theory \cite{brashers2003medical,brashers2004social,brashers2007theory,clayton2018theories,miller2014uncertainty,miller2020examining,middleton2012sources}.
By using UIT, the studies provide a comprehensive framework for viewing the experience of chronic illness. 

\subsubsection{Appraisal} 
Appraisal refers to the cognitive process through which individuals assess uncertainty and appraise it as either (a) danger or (b) opportunity. This assessment then guides the selection of coping strategies, affecting how individuals manage the uncertainties associated with their health. Successful adaptation to uncertainty involves aligning coping strategies with the individual's appraisal, ultimately shaping their ability to navigate and adjust to the challenges posed by the illness \cite{wright2009illness,mishel1981measurement,mishel1990reconceptualization,mishel1999uncertainty,wallace2005finding,clayton2018theories, mishel1988uncertainty}.

Health communication scholars have studied the relationship between a patient's appraisal of illness uncertainty and communication of information. Some studies particularly highlight the connection between a patient's evaluation of their illnesses and the communication of information regarding their health conditions. For example, a meta-analysis of the intersection between uncertainty and communication by Kuang and Wilson \cite{kuang2017meta} and Brashers's theory of communication and uncertainty management \cite{brashers2003medical,brashers2004social,brashers2007theory}, state that communication of information can be used to address patients' information seeking or avoidance of information where both are coping mechanisms of cognitive appraisal \cite{clayton2018theories}.  The magnitude of avoidance of new dissonance
is not influenced by the amount of existing dissonance and that spreading of
alternatives occurs before a choice. He proposes changing the definition of dissonance to include the degree to which a behavior will lead to a consequence
and the desirability of the consequence.

If uncertainty is appraised as ‘danger’, then one would adopt what Mishel termed `mobilizing strategies', like seeking new information or taking direct action. In addition to, or in conjunction with mobilization strategies, one might engage in affect-control strategies, like deciding to follow a religious faith or practice emotional disengagement, where they intentionally distance themselves from emotional involvement or investment in a situation, relationship, or experience to manage the emotional distress associated with uncertainty \cite{wright2009illness,mishel1981measurement,mishel1990reconceptualization,mishel1999uncertainty,wallace2005finding,clayton2018theories,mishel1988uncertainty}. 

Second, when uncertainty is appraised as ‘opportunity’, individuals may use `buffering strategies' as a way to block or neutralize new information or stimuli that could alter the perception of uncertainty from an opportunity to a danger \cite{wright2009illness,mishel1981measurement,mishel1990reconceptualization,mishel1999uncertainty,wallace2005finding,clayton2018theories, zhang2017uncertainty, mishel1988uncertainty}. This, in turn, would allow them to maintain the status quo and prolong the uncertainty. An example of perceiving uncertainty as opportunity could be a person who refrains from seeking a definite diagnosis or treatment for a better health condition. Suppose an individual is experiencing symptoms but decides not to undergo further medical treatment. A report on this phenomenon found that symptoms associated with medications were interpreted by HIV-positive individuals as evidence that medications were not working or had greater risks than benefits, which led to decreased adherence with HIV medications \cite{brashers2003medical}.

\subsubsection{Adaptation} 
Adaptation involves finding ways to live with uncertainty, adjust to changes, and maintain a sense of well-being. UIT emphasizes the dynamic nature of uncertainty, acknowledging that it can change over time as individuals gather more information, experience different aspects of the illness, and adapt to their circumstances. In some studies, adaptation is seen as a psychological and emotional adjustment to the challenges posed by the illness. It involves integrating uncertainties into one's life and finding a new sense of balance despite the ongoing health-related indecisiveness \cite{zhang2017uncertainty}.
Symptom appraisal in uncertainty with survivors of childhood cancer captures this for cancer patients in Heathcote's study, where one of her findings includes the topic of living with symptom-related uncertainty, which captured participants' recognition that post-cancer symptoms are tricky and influenced by psychological factors such as anxiety \cite{heathcote2021symptom}. Thus, positive adaptation doesn't necessarily imply the elimination of uncertainty but rather focuses on developing coping strategies that enhance well-being and overall adjustment despite ongoing uncertainties \cite{wright2009illness,mishel1981measurement,mishel1990reconceptualization,mishel1999uncertainty,mishel1988uncertainty}.

Individuals who positively adapt are able to integrate the uncertainties into their lives, find a new sense of balance, and adjust to the changes brought about by their health condition. Positive adaptations not only focus on managing uncertainty but also extend to fostering a sense of well-being and equilibrium despite ongoing health-related stresses. This adaptation is influenced by factors such as effective coping mechanisms, social support, and the individual's cognitive appraisal of uncertainty. Positive adaptation reflects resilience, the ability to maintain a positive outlook, and the capacity to navigate the uncertainties associated with illness without experiencing excessive distress. Earlier work shows that Malaysian breast cancer patients displayed elevated levels of uncertainty and were more inclined to employ avoidant coping (when uncertainty is appraised as opportunity) and less likely to utilize active emotional coping (when uncertainty is appraised as danger) \cite{sharif2017locus}. 

\begin{figure}
    \centering
    \includegraphics[width=1\linewidth]{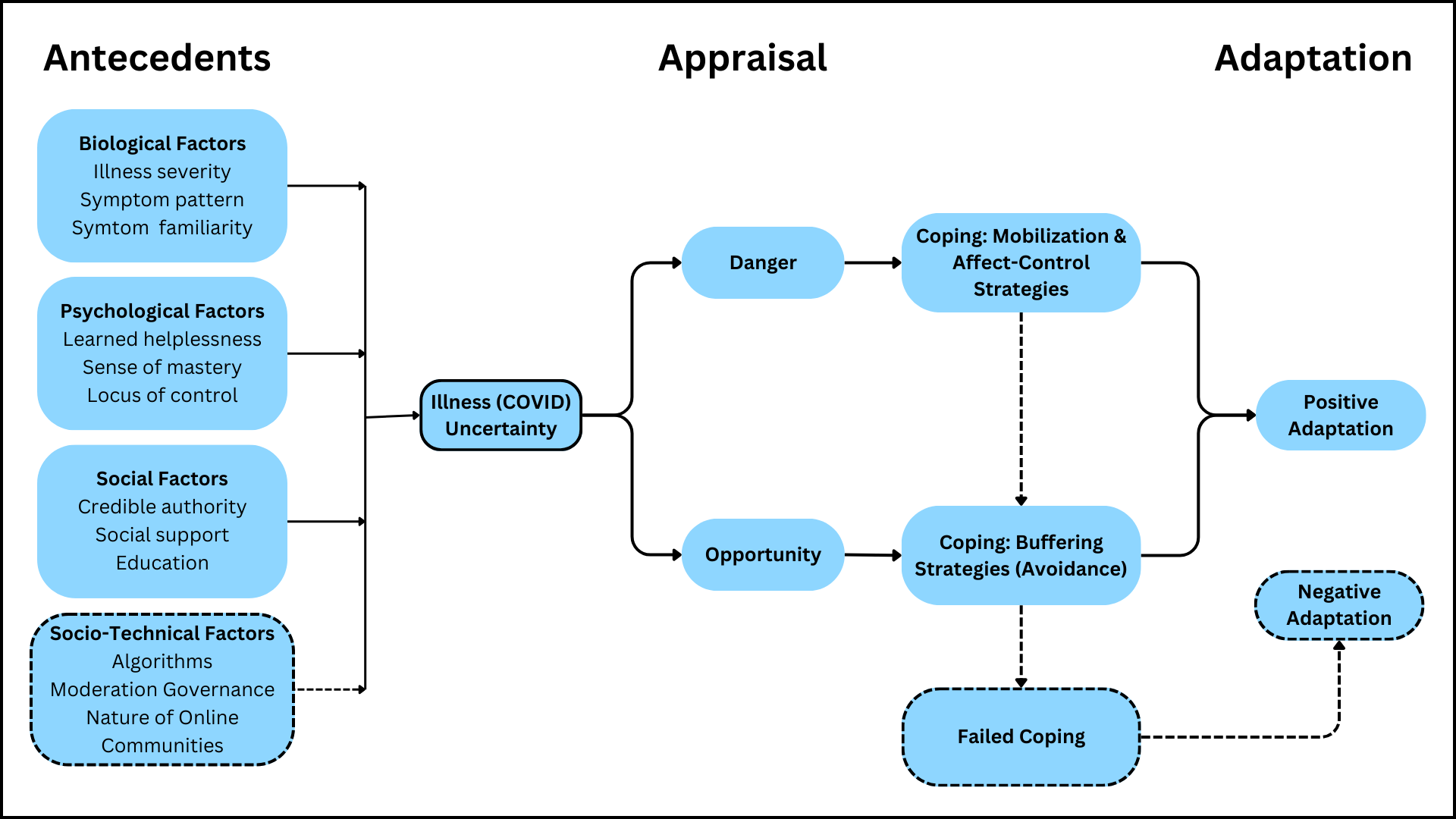}
    \caption{Theoretical model adapted from the Uncertainty in Illness Theory (Mishel, 1988) \cite{mishel1988uncertainty} and the review of Uncertainty in Illness Theory by Wright et al. (2009) \cite{wright2009illness} Our modifications to the model are highlighted with dotted outlines. We have introduced a new Antecedent: Socio-Technical Factors. In the Appraisal section, we demonstrate the impacts of this new Antecedent and introduce a component called `Failed Coping,' which is not present in the original theory. Additionally, we have incorporated a component in the Adaptation termed Negative Adaptation which incorporates `Burnout.'} 
    \label{fig:theory}
\end{figure}

\subsection{Finding parenting information and support online} \label{sec:parenting_related_work}
One of the earliest online communities - the Well, included a substantial amount of parenting discourse, ranging from the mundane (e.g., diaper changing) to discussions about teenagers and children with special needs \cite{rheingold_virtual_2000}.  On the Well's Parenting Conference, ``small but warmly human corner of cyberspace [provided] emotional support on a deeper level, parent to parent.'' Since then, many other parenting-centered platforms (e.g., BabyCenter \cite{beyers2022mother}, Mumsnet \cite{pedersen2018you}, YoubeMom \cite{schoenebeck2013secret}) and parenting online communities on major social media sites (e.g., Facebook groups \cite{lauren_et_al_19}, Reddit \cite{Sepahpour_Fard_mommit_daddit}, Twitter \cite{DeChoudhury_14, suh_et_al_14}, Youtube \cite{Borgos-Rodriguez_et_al_19}) have surfaced.  

Connecting with other parents in online communities gives parents social support \cite{plantin_parenthood_2009,lupton_parenting_2016,fleischmann_narratives_2004,lupton_parenting_2016,lupton_australian_2016}.  For example, mothers are empowered by and find a sense of community through blogging \cite{stavrositu_can_2008,madge_parenting_2006} and fathers introduced do-it-yoursef (DIY), a traditionally masculine activity to blog about their parenting experiences \cite{ammari_et_17}. Fathers, who have transitionally not been seen as caregivers, started looking online for other fathers facing similar challenges look for other fathers facing similar experiences and challenges to engage with online \cite{stgeorge_fathers_2011,salzmann-erikson_fathers_2013,ammari_schoenebec_fathers_15}. In fact, many stay-at-home fathers started creating online communities specifically built for fathers while excluding mothers \cite{ammari_schoenebeck_16}. This allows parents to seek information about their parenting experiences but also to make sense of their experiences \cite{lupton_parenting_2016}. 

\subsubsection{How parents make use of different platform affordances and signals} \label{sec:sub_parents_related_work_diff_affordance}
Social media platform affordances include association with anonymity and content recommendation \cite{faraj_materiality_2013,leonardi_social_2017}. Some social media platforms like Reddit afford users the capacity to post either anonymously (e.g., YoubeMom) or pseudonymously (e.g., Reddit). The anonymity affordance allows people to post more freely about stigmatizing and sensitive topics that might very well be difficult to share on Facebook due to context collapse \cite{hogan2013pseudonyms}. Many parenting topics, ranging from sleep training, vaccination, and postpartum depression, are considered sensitive. That might explain why many of these issues were discussed on Reddit parenting forums \cite{ammari_et_al_18}, while some of the more sensitive topics such as postpartum depression, were discussed using throwaway accounts - which allowed for even more anonymity \cite{ammari_et_al_19}. 

\subsubsection{Moderation of parenting spaces} \label{sec:parenting_related_work_moderation}
Volunteer moderators of parenting communities acted as gatekeepers, performing several crucial tasks to maintain the integrity and safety of these spaces. They (1) verified the identities of members to ensure authenticity; (2) accepted only new applicants who aligned with the group norms, fostering a supportive and like-minded community; (3) set the privacy policies of the parenting group, such as determining what information could be shared outside the group; and (4) removed members who violated these privacy rules \cite{mansour_et_al_21, li2024continue}. While these demanding tasks can sometimes be alleviated using automated moderation tools, these tools frequently fail to account for the nuances and complexities of online communities \cite{10.1145/3322640.3326699,gonccalves2023common}. Automated systems often lack the ability to interpret the subtle social dynamics and context-specific rules that human moderators navigate daily, leading to potential oversights and mismanagement. Thus, the role of  moderators remains indispensable despite the availability of technological aids. \cite{kuo_et_al_23, 10.1145/3322640.3326699,10.1145/3555111,gonccalves2023common}.

\subsubsection{COVID and parenting online}
The enactment of social distancing guidelines during the COVID pandemic led to an increased usage of social media \cite{cho2023bright}. People reported a surge in their social media activity compared to pre-pandemic times, driven not only by the availability of more time but also by the desire to stay socially connected and informed about pandemic-related news \cite{10.1145/3490632.3490666}. Since the onset of social distancing, both parents and their children have shown an uptick in their use of technology and social media \cite{drouin2020parents}, raising concerns about the potential effects of this increased usage \cite{10.1145/3416089, 10.1145/3386600}.

Richards et al. \cite{richards_et_al_22} and Matthews et al. \cite{matthews2022fathers} discussed how the shutdowns collapsed parents' professional and family lives by removing silos  (e.g., no school). Fathers attempted to limit the family's screen time and think about different ways to support their well-being \cite{matthews2022fathers}. However, parents with elevated anxiety levels were more likely to turn to social media for connection and information-seeking purposes \cite{drouin2020parents, 10.1145/3490632.3490666}. Prikhidko et al. highlight the risks associated with these actions for parents who may be susceptible to absorbing negative emotions encountered in their online activities, potentially leading to feelings of parental inadequacy, especially when compared to their pre-pandemic selves \cite{prikhidko2020effect}. Other studies have identified a positive correlation between COVID-related news consumption and increased levels of depression and anxiety in parents \cite{golding2021covid, 10.1145/3490632.3490666, whaley2023parental, prikhidko2020effect}.

Social media interactions have prompted individuals to adjust their behavior to safeguard their mental health. Heshmat and Neustaedter's findings on family and friend communication during the pandemic revealed a trend toward "technology detachment and cleanse" where some individuals became overwhelmed and chose to quit social media \cite{10.1145/3461778.3462022}. In other cases, people described self-censorship of their social media posts to avoid negative feedback from potential viewers \cite{10.1145/3490632.3490666}. Particularly among parents, some reported the need to exert control over the information they sought related to COVID to alleviate parenting difficulties \cite{adams2021parents}.

Despite the outlined risks that parents face on social media platforms, there are still benefits to the observed transformations in social media usage during the pandemic. In the healthcare domain, social media has played a critical role in disseminating COVID information and providing the general public with greater access to engage with government initiatives \cite{10.1145/3494825.3494830}. Various studies emphasize the maximization of efforts in utilizing social media platforms for public health by health professionals, including the risks and benefits to be considered \cite{chauhan2012social, lambert2012risk}.

\subsubsection{Parents navigating pediatric health information } \label{sec:parenting_related_work_pediatric health information}
Parents and caregivers are increasingly using the internet and social media for pediatric health information and parenting information \cite{bryan2020parental, wainstein2006use, pehora2015parents, plantin2009parenthood, duggan2015parents, 10.1145/3479546, drouin2020parents}. In today’s new post-pandemic world, the parental uncertainty that exists regarding pediatric COVID information is an under-studied space. Previous studies have identified variable accuracy of reliable health information across internet sources \cite{haddow2003caring, scullard2010googling, ammariparentspaper,adams2021parents}. Accessing healthcare information has become even more complex during the pandemic. Parents of children with medical complexities have been flooded with doubts about healthcare information based on what they read online \cite{nelson2010parenting, 10.1145/3479546, drouin2020parents, hood2021coronavirus, buonsenso2022caregivers}.

Despite the variable quality of information, it is found that only half of the parents re-verify the pediatric health information they find online with their practitioners \cite{ettel2012adolescents, pehora2015parents}. Some studies talk about parents and healthcare professionals sharing an inclination to investigate social media as a means of communication. Yet, there is a scarcity of substantial evidence to endorse optimal approaches or practices for speaking about healthcare in an online space \cite{bryan2020parental, radford2016volunteer, li2022channels,knight2005developing,lonzer2015social,delago2018qualitative}. 

Now, this journey of parenthood in a pandemic-laden world intensifies uncertainty, creating an environment where stress and anxiety become inevitable companions. As parents wrestle with social distancing and quarantine protocols themselves, the uncertainty surrounding their roles and responsibilities as new parents or parents of CMC (children with medical complexities) compounds \cite{nelson2010parenting, goldberg2021parenting, 10.1145/3479546, drouin2020parents, hood2021coronavirus,cohen2011children}. It is crucial to recognize that the uncertainty experienced by parents, especially new parents, lacks any historical precedent that could serve as a guiding reference for managing such profound uncertainty with their children's or their own diagnosis of COVID as an illness \cite{hood2021coronavirus}.

Parents whose children received behavioral care, special education, or any other kind of support found themselves having to take the place of the professional because their children could not get ``that special help they need'' after the lock downs \cite{richards_et_al_22}[P.11]. To support their children, parents had to access information online and coordinate with professionals possessing the knowledge and the experience. The authors argue for ``scaffolding parent information seeking by validating sources, sharing personal sources, or facilitating peer support'' \cite{richards_et_al_22}[P.21]. This scaffolding is important because parenting information is usually highly contextualized and difficult to apply to all family situations \cite{ibrahim_getting_advice_24} even if the ``steps'' parents need to take are known - the authors refer to this as the information-to-application gap [P.17]. Indeed, parents use``real-world'' advice from trusted others when they are being ``bombarded by all this information.'' \cite{Kirchner_et_al_20} Kirshner et al., referred to this process as the parents' intuition \cite{Kirchner_et_al_20}. One of the ways they suggest information be provided to parents is through story-telling as it can help technology be more socially oriented and attuned to the needs of the parent.

The unprecedented nature of the situation underscores the immense challenge faced by individuals in navigating the uncharted territory of parenthood amidst a global pandemic. In this uncharted landscape/environment, the emotions of uncertainty, stress, and anxiety intertwine, forming a complex tapestry that requires understanding, support, and resilience from parents and the community alike \cite{nelson2010parenting, goldberg2021parenting, 10.1145/3479546, drouin2020parents, hood2021coronavirus}. For example, parents of children with special needs look for case-specific (e.g., autistic children) and localized groups (e.g., parents of autistic children in New York) to find parents who face similar challenges and who deal with similar uncertainties which in many ways are contextualized by their location. When negotiating the Individualized Educational Plan (IEP) with their school, parents try to find others who have negotiated the plan within the same school district to have a better sense of what services they can provide their child \cite{ammarinetworkempower}.

\section{Methods}
In this section, we first start by describing our data collection process including recruitment from different sources. We then describe the qualitative analysis methodology used in our analysis. 
\subsection{Data Collection}
We interviewed 23 birth parents using semi-structured interviews. 
Participants were eligible for this study if (1) They were over the age of 18, (2) They were a parent to a child or children who tested positive for COVID and (3) They were in the United States. There was a 20\$ gift card compensation for successfully completing the interview.

Participants were asked first to complete a Qualtrics questionnaire that captured their demographic information, resident state of the US, number of children, children demographics, age at which the child tested positive, social media sites or online websites they use to find information, highest completed level of education, and the approximate household income. We deployed this survey on Twitter, Reddit, and Facebook. For Facebook and Reddit groups, we asked permission from the group administrators before posting a recruitment message. 

In this first stage of recruitment, we came across large instances of bots and fake participants. Recent studies call for adaptive strategies to deal with such scenarios. We adhered to the guidelines for ethical research \cite{10.1145/3613904.3642732, aluwihare2012ethics, 10.1145/3411764.3445584} to make our own specific small-scale criteria to prevent and handle fraudulence.
First, we clearly identified that we encountered subjects who did not reside in the United States of America yet were still signing up for the survey. Our initial survey was a total of 874 responses. 
We cleaned the initial response list using three data-cleaning techniques:
\begin{enumerate}
    \item We excluded respondents that had multiple responses by the same IP addresses. A frequent occurrence we observed was that numerous distinguishable responses originated from the same IP address since this would make it possible for one person to qualify for the interview multiple times by impersonating different people which then qualified for the compensation multiple times.
    \item We matched the IP addresses to the location they filled out on the survey. We found that fake subjects claimed to be from a state but their geo-location was outside of the US or in a different state. We accounted for participants who may have used a VPN (virtual private network) by carrying out the next step.
    \item Qualtrics surveys have a unique feature of recording the approximate latitude and longitude of the device used to fill the survey. It does not pinpoint a location but approximately displays a potential state in the US where the survey might have been answered. We cross-referenced this to the state where the participant claimed to reside. 
\end{enumerate}

After implementing these three cleaning steps, there remained only 250 unique participant responses. Using a purposive sampling technique, we asked 100 participants to sign up for the semi-structured interview at their time of convenience. Thirteen parents responded in our first round of recruitment. We decided to have a second phase of recruitment because we did not reach data saturation. We posted the study on ‘Volunteer Science' \cite{radford2016volunteer}, a research recruitment platform that provides a pool of online volunteers. 

We also employed an in-person method of recruitment where a new IRB was submitted to recruit participants in the waiting room of a local teaching hospital. Flyers were placed at the reception after gaining permission from the department officials in the pediatrics specialty, and pasted on notice boards in the waiting areas where patients could scan the QR code to demonstrate an interest in the study. 

Additionally, two pediatricians helped us recruit respondents whose children were `COVID admissions' in the hospital and, via email, sent them information about our study. If the parents were interested, they would fill out the questionnaire, and then the first author would reach out to them to schedule a time for the interview. We recruited 8 new interviewees in this phase of recruitment.

All of our 23 participants had used social media at some point to look for COVID-related information for their child/children. All interviewees and their children tested positive for COVID. All interviewees held at least a four-year college degree. Eleven participants had bachelor's degrees, eleven held graduate degrees, and one held a doctorate. We show the participant details in Table \ref{tab:participants}.

All participants held at least a four-year college degree. Thus, the study focuses on college-educated parents in the USA. We had only one participant who fell under the low socioeconomic status (SES) category.
However, this individual had a college education, which indicates a level of upward mobility. Studies suggest that better socio-economic situations allow parents to have ‘improved coping’ with challenges that bring various uncertainties of everyday life and parenting \cite{mascheroni2018digital,scrimin2022effects}. Additionally, college education significantly influences individuals' ability to navigate and critically assess information, especially in the face of misinformation and health information \cite{hwang2023education, hayward2023dynamic}. 

A myriad of studies significantly support the idea that those with lower education were more likely to accept misinformation. The impact of education on misinformation acceptance is mediated by various media dependencies \cite{hwang2023education,gaziano1997forecast,sun2022battle}. We found that parents were inclined to verify facts from multiple channels, cross-reference information with Google Scholar, and rely on discourses on online platforms to formulate strategies to deal with their uncertainty. We have explored these further in the findings.

\begin{table}[h!]
\begin{tabular}{cccccl}
\textbf{Participant} & \textbf{Gender} & \textbf{\# Children} & \textbf{Child Gender} & \textbf{Child Age} & \multicolumn{1}{c}{\textbf{Social Media Use}} \\
P1                        & F                           & 2                        & F, F                  & 18, 12             & FB                                            \\
P2                        & F                           & 1                        & F                     & 9                  & FB,LI, IG                                     \\
P3                        & F                           & 1                        & M                     & 7                  & FB                                            \\
P4                        & F                           & 2                        & M, F                  & 5, 1.5             & FB, WA                                        \\
P5                        & F                           & 1                        & F                     & 18 months          & WA, IG                                        \\
P6                        & F                           & 1                        & M                     & 17 months          & WA, FB, IG, RT, PI                            \\
P7                        & F                           & 1                        & F                     & 14 months          & WA, FB, IG, ST, TT                            \\
P8                        & F                           & 1                        & F                     & 22 months          & WA, IG                                        \\
P9                        & F                           & 1                        & F                     & 18 months          & IG, RT, WA                                    \\
P10                       & F                           & 1                        & F                     & 8.5 months         & TT                                            \\
P11                       & F                           & 2                        & M, M                  & 2, 2 months        & FB, TT, IG                                    \\
P12                       & F                           & 2                        & F, F                  & 4, 1.5             & FB, IG, X, RT, LI, TT                         \\
P13                       & M                           & 1                        & M                     & 11                 & FB, IG, X, YT                           \\
P14                       & F                           & 2                        & M, M                  & 6, 3               & ST, IG, TT                                    \\
P15                       & F                           & 1                        & F                     & 13                 & DD, RT                                       \\
P16                       & F                           & 2                        & F, M                  & 12, 11             & FB, IG, X, Y                                  \\
P17                       & M                           & 2                        & M, M                  & 31, 21             & FB, X, IG                                     \\
P18                       & F                           & 2                        & M, F                  & 21, 18             & X, FB, IG, LI, ST                             \\
P19                       & M                           & 1                        & M                     & 3                  & LI                                            \\
P20                       & M                           & 1                        & M                     & 3                  & LI, FB, IG, ST, TM, WA                        \\
P21                       & F                           & 3                        & F, F, M               & 17, 15, 10         & FB                                            \\
P22                       & F                           & 1                        & M                     & 19                 & FB                                            \\
P23                       & M                           & 2                        & F, F                  & 11,14              & FB, X, IG, LI                                
\end{tabular}
    \caption{\label{tab:participants} This table lists our participants along with their self-reported gender. We also report the number of children, their age (in years), and their gender. This table shows the social media sites used by each participant. Social media platforms are abbreviated. FB: Facebook; X: X, formerly Twitter; IG: Instagram; RT: Reddit; LI: LinkedIn; YT: Youtube; ST: SnapChat; TT: TikTok; WA:WhatsApp; PI: Pinterest; TM: Telegram; Discord: DD}
\end{table}

\subsection{Data Analysis}
All interviews were conducted by the first author (n=23). Interviews were transcribed using Otter AI. The transcription process included replacing all identifying information with pseudonyms or place markers such as [older child], [younger child], etc., after which they were imported into NVivo. The interview questions are shared in the Appendix (Section \ref{sec:appendix_interviews}).

Sampling concluded at 23 interviews due to thematic saturation, where recurring themes and patterns emerged consistently across interviews, suggesting that further interviews within the same sample constraints would unlikely yield significantly new insights. For example, participants frequently expressed reduced anxiety about COVID vaccination for their children after observing positive experiences shared by others on online platforms. Another recurring theme highlighting data saturation was parents consistently selecting online groups they deemed suitable for their children based on their own judgments. Notably, as mentioned in our findings, there was a deviation from their personal ideologies to join groups recommended by medical professionals, such as WhatsApp ``bumper'' groups created by practitioners specifically for mothers. This resulted in a challenging cycle where parents engaged online but avoided spaces that conflicted with their beliefs. When questioned about this, they explained their strategies for identifying “credible sources” and navigating COVID uncertainties, which we also detailed in our findings.

We first reviewed four interviews selected at random individually and discussed key themes. The themes were added to a code book containing a description of the code and an example quote. This code book was created to standardize the analysis process before it started. Given that this study involved semi-structured interviews, a standardized code book with iterative discussion helped ensure that all themes were brought to light. This also helped to exclude some existing well-studied literature like child sleeping patterns, feeding habits, etc. which were not the primary objectives of this study.
Code definitions and examples were decided mutually in the weekly meetings until a consensus was reached for each code \cite{McDonald_et_al_19}. Then, the first author coded all interviews and applied axial coding to explore relationships using the code book \cite{strauss1998basics, sofaer1999qualitative}. Each coded sheet was then reviewed and discussed by the group iteratively  \cite{sofaer1999qualitative, neale2016iterative}. The discussions eventually brought to light recurrent topics and implications that we explore further in the findings. The produced code book is shared in the Appendix (Section \ref{sec:appendix_codebook}).

Any disagreements on the recurrent topics and implications were discussed, reaching consensus (as described in \cite{McDonald_et_al_19}) during weekly group meetings among all researchers by rigorously adhering to the guidelines outlined for reliability in qualitative research \cite{McDonald_et_al_19,neale2016iterative}. In line with earlier works in CSCW  and CHI, since the emergence of our themes is part of the coding process, we did not calculate the inter-rater reliability for codes \cite{McDonald_et_al_19, braun2021thematic}. This was a consensual decision by the authors as thematic analysis for semi-structured interviews can be vast and rich \cite{braun2021thematic}. The goal of our analysis was to identify emerging themes in our data rather than capture the frequency of predetermined attitudes or common occurrences. 
\cite{McDonald_et_al_19, braun2021thematic}.

\subsection{Researcher Positionality}
The concept of leveraging our innate instincts, developed over time through evolution, to enhance different aspects of our lives is intriguing. These instincts, crucial for our survival and prosperity throughout history, can be utilized in contemporary settings to improve our well-being, family relationships, and professional performance \cite{weisinger2009genius,bornstein2005handbook}. 

The first author, a South Asian first-generation U.S. college student and resident brings a unique perspective to this research. With a background in UX research and design and as the descendant of two physicians, has done research that focuses on understanding technologically mediated communication in healthcare for marginalized populations and collaborative healthcare teams.

The second author on this paper, an East Asian who is a first year med-student with the goal of conducting research on the experiences of marginalized communities in the global healthcare system.

The third author, was raised and educated in the Middle East, brings a notable cultural perspective to his research. His work has focused on the use of technology by members of marginalized groups in both the Global South and the Global North.

These unique perspectives drive the exploration of how parental instincts in navigating uncertainties regarding children are being amplified and have become uniquely distinguishable in modern contexts, particularly in the wake of various pandemics, with COVID-19 being the prevalent one.

\section{Findings}

The era of COVID has introduced unprecedented challenges, uncertainties, and transformations. This global pandemic disrupted daily life, altered social norms, and changed how people work, interact, and perceive the world. The implementation of new health protocols, lockdowns, and the rapid dissemination of information, sometimes misinformation, has heightened a sense of unpredictability, especially in parents. The repercussions on mental health, economic stability, and societal dynamics have created a uniquely surreal environment, making it an extraordinary and strange period in recent history. 

Within this context, being a parent or becoming a new parent has become an even more challenging experience. Navigating parenthood in a pandemic world characterized by masked faces, social distancing, and a collective sense of uncertainty has placed parents in unfamiliar and unpredictable situations.

All twenty-three parents we interviewed for this study described being uncertain and overwhelmed by the changes brought about by the pandemic. Social isolation due to lockdowns and other COVID restrictions added to the uncertainty they were all facing with a novel virus. This uncertainty applied to both first-time parents and parents expanding their families by welcoming a new addition.

P8, a recent first-time mother, says  \textit{“We had [the baby] during COVID. So there was no pre-COVID knowledge of how to raise a child for us…It was kind of a weird situation, like, do you take her out? Do you not take her out? What's-- she can't be vaccinated till she's three months old. So it was like a weird time to have a kid. Do you mask? Do you not mask? How do you-- you can't mask a baby. How do you-- it was-- you just really had no idea what to do. There were no rules. And looking back on it, there was no right or wrong way. Honestly, it's just trust your gut. But yeah, we were just new parents in like a very strange world.”} 
\newline This uncertainty was not limited to first-time parents, however. P4, whose child was infected with COVID said that she \textit{“kind of knew what his flus were like, his ear infections. But this was different. This was 100\% different. The symptoms were way more severe. They persisted for much longer…it hit us…this is not normally the case”} with their child.

P11, among other parents, also felt alone in the postpartum period and \textit{“just [wanted] to be heard and understood.” }She turned to social media where \textit{“people were posting about having a new baby and how challenging it is and I have a new baby and it's challenging.”} 
Having these shared experiences with other parents who are themselves feeling alone during the time following childbirth was\textit{ “very validating”} to P11. In the next section, we expand on the different ways social media provided a positive space for caregivers in the post-COVID period. We also show examples of when online communities fell short in supporting parents. 

\subsection{Parents establish a networked understanding of COVID} \label{sec:networked}
As we introduced in Section \ref{sec:parenting_related_work_pediatric health information}, parents look for child healthcare information in their social networks and through online sources. In this section, our findings expand on this earlier work to show how parents come to grips with COVID by adapting to the uncertainties brought about by the pandemic, including many changes in the overall well-being of their families. Parents looked for different ways to navigate concerns related to childcare and addressing the potential social and emotional effects of disrupted routines. Coming to grips with COVID as a parent involved a blend of practical adjustments, emotional support seeking, and a commitment to staying informed about health guidelines. The journey for parents during this time required resilience, adaptability, and a focus on maintaining the overall well-being of the family unit in the face of unprecedented challenges.

\subsection{Medical Professionals and Social Media: Medical Authority vs. Parental Community}
In keeping with the earlier points about COVID being an unprecedented event in the parents’ lives, P8 expressed that \textit{“even each [different] pediatrician would tell us something different. Parents are unsure of how to have them interact until they get all of their shots. We had like an added layer of COVID, like how do you navigate that really? So that was more of just as if there's no rules. No one could tell us what to do. Everyone was telling us something different. Every pediatrician would tell us something different.” }

\subsubsection{Parental strategies to find trusted medical information in uncertain times}
The divergence in pediatrician views motivated parents to start finding online communities or medical sources on the web which can shed some light on the issue. 
P5 said that to her \textit{“Instagram like, before the pandemic was really just for friends. And taking pics, whatever. It was more social. But I would say Instagram for me now is much more information-driven.”} 

P12 echoed this sentiment saying that the \textit{“evidence-based information”} that she found through Facebook was \textit{“super helpful”} to her as a new parent. The participants indicated that \textit{“evidence-based information”} was knowledge shared within the online community from the real-life experiences of other parents with their children. P12 explained further that from such evidence-based information, she  \textit{“learned from others who have gone through it … before that was during the pandemic I was not really really so active member of this Facebook community…[because it] connects several women together. Where we can get resources also..or just someone we can talk to - someone who can listen because [when] you wake up a morning and your children are having flu-like symptoms….So you ask other mothers….The Facebook groups have [this]-- Facebook group helps me a lot.”} P4, whose toddler \textit{“kept throwing his mask on the floor”} found the resource she needed online when parents on a Facebook group suggested that they not \textit{“get the paper ones because they chew right through them. Put on a cloth one, or have him wear his favorite color mask or a favorite cartoon."} P4’s child got a mask with those properties which he was willing to wear because she found that idea through parents in the same situation. 

The content within these online communities being (a) referenced information (with clearly cited sources) and/or (b)  experience-based  is what makes chats, comments, etc. shared essential in promoting the use of these communities among parents. Parents relied on other trusted parents' testimonials for discussions about their own children’s health and well-being. We found that the term "evidence-based information" carries varied interpretations among parents. For some, it encompasses reassurance from other parents' lived experiences, while for others, it implies information shared within the online community should be substantiated with sources linked to reputable web pages such as the CDC, Mayo Clinic, or even a pediatrician’s personal web page. For parents, it is not merely a suggestion for solving a particular problem; rather, it involves finding a practical approach from someone who has explicitly experienced their concern or uncertainty. That someone is usually someone they trust, or with whom they have a lot in common.

There is a duality to parents' understanding of what constitutes `evidence-based' information because it is often anecdotal evidence. Our study identified that parents' concept of evidence-based information encompasses two key, yet contrasting components:

\begin{enumerate}
    \item \textit{Information provenance}: Parents demonstrated their commitment to evidence-based information by sourcing it from reputable academic and medical resources. For example, P14, referred to as the “Vanderbilt mom,” specifically used resources from Vanderbilt University. Other parents sought out literature on Google Scholar, thereby grounding their approach in solid, evidence-based research. Additionally, P4 found mom groups on Facebook and WhatsApp particularly helpful for local updates and activity recommendations during lockdowns. However, she was cautious about fully trusting anecdotal experiences shared on forums, preferring more generalized and scientific information from reputable sources. She extensively used online resources like Google, the CDC website, and news articles to research symptoms and verify information.
    \item \textit{Community trust and outsourcing research}: The second component of "evidence-based" information is the trust parents place in their communities, which they often rely on to `outsource' research. These trusted communities, both online and offline, played a crucial role in providing invaluable support. They shared cases of similar health conditions and directed parents to relevant resources, whether from other online sources or other medical practitioners. This collective approach helped parents gather comprehensive and reliable information about their child’s health condition. When parents rely on their communities, the community effectively becomes a scaffold, allowing parents to delegate some of the mental burden. This shared responsibility helps parents manage the extensive information required to care for their children. These insights of what parents consider 'evidence-based' emphasize the importance of reputable information sources and community trust in parental decision-making.
\end{enumerate}

P22 is one such parent who also stated that she \textit{“found the comments ‘enriching’”} and trusted Facebook groups’ content because of moderators who managed the Facebook group  because group members were not allowed to post unverified information about CF treatments. If any posts are deemed unacceptable, the moderator \textit{“will like, erase your comment. It's just very professional.”} Members who frequently offended were blocked from the group. P15 shared that they also accessed this information through their social network because \textit{“usually the people that I'm getting information from online… they have sources and they're not giving me opinions so much as sharing the information that they found.  So that you can then follow the information that they've found and make your own judgment….”} P6 echoes this sentiment saying that verifying the credibility of information on Facebook or Reddit - \textit{“and it goes back to the credibility [of the social media post] .. if you can find that information on the CDC or a couple of the other more legitimate websites or if you do a quick Google Scholar search where it’s there then all right, that's a thing. I'm more-- I'm much more inclined to believe it versus where it's just kind of said and you can't really find it replicated in other places.”}

\subsubsection{Parental strategies for verifying credible sources}
Parents reflected on how during the early stages of the pandemic there were challenges faced due to the fluctuating nature of information flow. The lack of centralized guidelines and real-time insights on COVID treatments online, particularly during critical situations like the surge in COVID cases observed in places like New York, Italy, etc., posed significant difficulties for parents as well as healthcare professionals. Parents like P18 thought that \textit{“it'd be nice if there was like 1 central one that did everything instead of you having to go multiple sources to see if it's in your county and your state in your country. It felt like there were ten different [online] places you had to go instead of just one central place.”}

They saw this constant effort to follow the ever-evolving science as a privilege for the well-educated. When using search strategies, the information that has been cited and sourced to websites with domains like ".org" or ".edu" indicated to parents as more credible and trusted sources according to five of our respondents. 
P7 shared with us how the website domain determines if the information can be trusted, \textit{"Is it a .com? .coms in my brain still can't be fully trusted, because anybody can make them. So I sort of fall back onto information like Do they actually have relevant experience? Or are they just some random person saying whatever they want?"}

Three of them favored links affiliated with specific medical universities or hospitals. Eleven of our participants mentioned that they use Google Scholar actively. 

Parents also mentioned that trust in information was not just through online communities and testimonials. It was also from parental instincts about what they found did not appear as spam or was tagged as ‘sponsored content’. P23  highlights how \textit{“Wikipedia…is a [community]. It is possible for basically anybody to make an account and make new information … But it’s a motivated, driven community... if they see something that is poorly researched or factually incorrect, they're empowered to wipe that out and say no - here's the actual truth of this... And here's my sources.”} This is a deliberate choice made by caregivers that involves seeking information exclusively from selected groups while actively avoiding exposure to everything else.

\subsubsection{How parents see moderation of medical information in their online spaces} 
\begin{figure}
    \centering
    \includegraphics[width=0.25\linewidth]{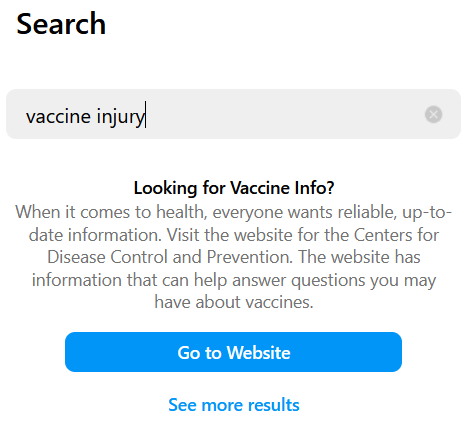}
    \caption{This is the message that appears to Instagram users who try to search for COVID vaccine injury. The link takes the user to the CDC vaccines and immunization page.}
    \label{fig:vax_instagram}
\end{figure}

As we outlined in Section \ref{sec:parenting_related_work_moderation}, moderation decisions influence the type of discourse that is acceptable in parenting online spaces. Access to information can also be restricted as a moderation decision on some platforms. For example, P9, a new mother who identified herself as an \textit{``anti-vaxxer"}, shared that she obtained information on topics of prenatal and post-natal care such as pelvic floor PT, prenatal chiropractic, birthing classes, and breastfeeding through Instagram. She mentioned using the platform to learn about these aspects, connecting with an online International Board Certified Lactation Consultant (IBCLC) for a virtual session to address breastfeeding concerns. However, P9 noted that Instagram is becoming more difficult to use for  \textit{“information curation.”} She did not want to take the vaccine and was against her child taking it either, specifically because she was worried about vaccine injury. She continues by saying that when attempting to search for vaccine injury on Instagram, it does not appear to be a \textit{“valid hashtag.”} Instead, she says that Instagram \textit{“flags [vaccine injury-related content]. It doesn't even let you search it. It says like, `You want to learn more about vaccines? Click here.’ There's limited information on stuff that doesn't fit the narrative that they want you to follow.”} Instagram is one of the social media sites that did not allow for posting about self-reported vaccine injuries with a clear message transporting the user to the CDC vaccines and immunization page.\footnote{\url{https://www.cdc.gov/vaccines/index.html}} As shown in Figure \ref{fig:vax_instagram}, you will see a message that takes the user to the CDC website. Vaccine injury refers to adverse effects or harm caused by vaccines. It includes any significant harmful reaction or side effect resulting from the administration of a vaccine. In P9’s case, she is referring to the vaccine injuries caused by COVID.

Conversely, P4 found it difficult to trust the information on social media for the diametrically opposite reason, specifically the lack of moderation since the content is \textit{“all anecdotal experience where people tell you what they feel…I try more so to rely on if I can find more scientific information, something from a more reliable source, something from like a government-- not necessarily government agency but something CDC-ish.”} Similarly, P14  called herself a Vanderbilt mom because she only trusted information from the Vanderbilt website - the same university her pediatrician graduated from. Additionally, when wanting more context about an online medical source, P3 found it difficult to contact her pediatrician to say \textit{"Oh, I just read this, because they're not gonna get back to you that fast.”} 

\subsubsection{Medical professionals as online community creators}
Health professionals also took on an active role in building supportive online communities. P6 explained how she joined the WhatsApp “bumper” group experience when \textit{“we [expecting mothers] were brought together by [our doctor at the time] when we were all still pregnant. So it was like a group of women who would give birth within a certain period of time.”} P6 explained further that  this \textit{“WhatsApp group for [parents of] babies born in the same range of months became the primary source of information”} for her. 
 
Twelve of our respondents mentioned being part of similar WhatsApp groups. Four of them saw information about our study on said WhatsApp groups. These WhatsApp groups were created by hospital staff where the participants received prenatal care; however, the doctors or the nurses did not actively participate in these groups. It was strictly mothers who were going through the same trimester, or were only at the most a couple of months apart in giving birth.

Participants found their information needs were being met by evidence-based information on these “bumper” groups. Experiential information from other parents who might be facing similar challenges is also paramount. For example, P8 says that whenever she has any questions about what to do for her child, \textit{“I'll go ask this [WhatsApp Group] first. You know, it was like, even when [my child] had her symptoms that I thought were COVID, I kind of asked them what their thoughts were. Should-- when-- how do I get her tested? What's the process of getting her tested? Or should I wait to see if this dies down? If it's just a cold? How do you know?!”} One concrete example of the importance of experiential information when filling the gaps that might be left by professionals comes from P8, whose doctor forgot to mention the key detail that both she and her husband needed to produce a negative COVID test to be allowed into the delivery room on the day of the delivery. She \textit{“found this out through the WhatsApp group when it was mentioned by another expecting mother. I guess [that’s why I have] trust in these people because they were sharing their personal information with me to help me.”} 

\subsection{How Parents Navigated the Pandemic Landscape: Parenting Online Communities, Lockdown Realities, and COVID Infections}
In Section \ref{sec:parenting_related_work}, we reviewed earlier work on how parents use social media platforms to make sense of their new identities as parents. Parents use different social media platforms depending on their respective affordances (see Section \ref{sec:sub_parents_related_work_diff_affordance}) to access support for special needs (e.g., disability or chronic conditions), social identity (e.g., mothers only or fathers only groups) and specific locations (e.g., neighborhood). The limitations on physical interactions, social gatherings, and support networks have created a void that intensifies the challenges posed by the uncertainty surrounding the pandemic. The emotional toll of navigating these uncertainties without the usual avenues of social support further highlights the importance of fostering alternative means of connection (e.g., in online communities) and support during these challenging times.

In addition to the “mothers only” group, we found another category of WhatsApp groups which were "hyper-localized" groups. These specific groups offered information pertaining to events and developments within the immediate neighborhood and residential areas surrounding the location where the parents resided. Some parents created WhatsApp groups that were analogous to their neighborhood communities. This allowed WhatsApp groups to be proxies for socializing at daycares, parks, and other physical third spaces - many of which were locked down early on in the pandemic. P4 explains that \textit{“[WhatsApp] was a great information source in terms of what's happening in the neighborhood, especially when we were all locked down and we had nowhere to go.”} These groups became “crowd locators” for parents to identify areas with mass gatherings, enabling them to avoid such places \textit{“we were trying to socially distance.”} So, if everyone is going to a public space in their neighborhood to watch fireworks, \textit{“I would say [to the kids] that we’re not going [there] because it sounds like everybody's going there.”} 

When P4 wanted to get a mask with a cartoon character for her toddler,  she \textit{“found a woman nearby, actually, again, through social media. She was making masks for kids. And she had like different fabrics and patterns and stuff like that. And she was selling one that was Batman.”} P4 and other toddler moms also said that they used \textit{“daycare [WhatsApp] groups”} regularly where they \textit{“would talk to the different daycare moms”} about their child’s health, specifically measuring COVID infections in daycare.  

As more people became infected with COVID, at times, parents found that seeing other people’s experiences with COVID on their social media sites reduced their uncertainty.  They thought that most were recovering from their infections, especially after vaccinations and antiviral medications were being introduced. P1 said that posts from people about COVID \textit{“treatment guidelines, it made me less anxious. Back in May… we were all sick, and when (younger child) got sick again in September, definitely made me less anxious just because I had that additional knowledge that for the most part, people are doing well and not having serious consequences versus back in 2020 or 2021.”}

As a self-professed COVID-cautious parent, P15 curated her own online community and content on Reddit by joining those subreddits where she found people willing to share resources and information about COVID mitigations (e.g., where to get good masks) that she could trust. P15 joined the r/ZeroCovidCommunity\footnote{\url{https://www.reddit.com/r/ZeroCovidCommunity/}} which is defined as \textit{“a place where people who are passionate about reducing the transmission of SARS-CoV-2 and protecting those who are at high risk can come together with a shared vision for the future. We are here to discuss the pandemic and politics, share info, and support one another. Individual mitigations can be a forced response to a lack of systemic solutions, but it is the latter that should be the main subject of activism. Engineering is better than hate.”}  She found Reddit communities engaging in supportive comments which gave her a sense of visibility and acknowledgment, contributing to the feeling of being seen and heard as a COVID-conscious parent even when most people around her did not engage in COVID mitigation strategies any longer. P15 not only received information and support from other users of the subreddit, but she was not \textit{“just taking from these communities, I'm also giving back and so it helps to remind you know, you’re not alone.”}


\subsubsection{Parental Perspectives: Managing Pre-existing Medical Conditions and the Role of Social Media}
Nine of our respondents were parents to children who had medical complexities independent of COVID. One of the parents, P2 \textit{“was hoping that I could find a[n online] parent support group”} to support her in her parenting challenges that were made more complex by her child’s medical condition, Strabismus, which is a \textit{“misalignment of the eyes, causing one eye to deviate inward (esotropia) toward the nose, or outward (exotropia), while the other eye remains focused.”}\footnote{\url{https://www.hopkinsmedicine.org/health/conditions-and-diseases}} While this misalignment can occur in children at an early age, it should not persist after the age of three months. P2 believes her daughter who had eye surgery for strabismus at 4 (but now was 9 years old) contracted COVID from the emergency room she was taken to after receiving a concussion from a fall at her school. P2 voiced to us her concerns of how after COVID and the concussion, her child started complaining about double vision again. P2 wondered if there was any correlation between COVID, the Pfizer vaccine, and strabismus. This pushed her to look for answers from other parents who have experiences like her own “single-mom family” online. She noted that the need for more specialized medical services was compounded by COVID infection risks and restrictions. P2 concludes this experience by saying that she felt the need to find \textit{“parents who have had kids [needing] strabismus eye surgery [at the time]. And I haven't found that…at the time, I was thinking, "Oh, my God, I could use some support.”}

P22 on the other hand, found that supportive online space that eluded P2. As a parent of a child with cystic fibrosis (CF), P22 is an immigrant mother who moved to the US right after the US FDA approval for ‘Trikafta’, a special drug for CF patients that increases life expectancy drastically. The drug was announced in October 2019 right before the pandemic. Her son who is a teenager contracted COVID after migrating to the US but being a CF patient is always uncertain and on high alert about his surroundings. She shared with us how \textit{“[Trikafta] fixed his lung issues. He doesn't have a lot of infections anymore. Of course, he still has to be careful with COVID. And with the flu season now it's like the alarm is always there in [our] mind, whenever he goes-- when he goes in class, for example, he's always like paying attention... Who's coughing, who's sneezing? He tries to be careful as much as he can.”}

She also spoke to us about ‘Paxlovid’, which is a medication used for the treatment of COVID-19. It is an oral antiviral drug developed by Pfizer for the treatment of mild to moderate COVID-19 in certain high-risk individuals. P22 expresses concerns about Paxlovid taken by CF patients, citing negative experiences shared by some people on the Facebook group. Initially hesitant, P22 became more open to the idea after hearing from her trusted online community that she called \textit{“a responsible group”} -  that it was a necessary medication for the protection of her son. P22 highlighted the importance of two sources of information in the Facebook Group. One was the posts  made by veteran patients and caregivers. The other was posts and inputs in the form of comments made by healthcare professionals who were also members of the group. Both of these emphasized the individualized nature of treatments for different patients. Having access to earlier similar experiences helped her overall uncertainty when it comes to augmenting COVID with other diseases. Given the complexities of caring for a  \textit{“son with a chronic disease…you feel alone sometimes…you feel like [you’re] isolated. The doctor…OK…will give you his time, but the doctor Is busy, so you feel…in this group, they would give you a lot of their experience. And sometimes a parent’s experience is very helpful because they've been through it all.”} 

P22 talks about how prior to her son contracting COVID, while having the uncertainty of how COVID affects CF patients, and while being in a country outside of the United States, where access to hospitals and doctors was limited and \textit{“knew less about cystic fibrosis than in the US”}, she turned to social media, specifically to a Facebook page - \textit{“This [group] did help me for a period of time, especially when I was not living in the US. So this group is mainly in the US so you get to understand the care that they receive here and you get to learn from it. And sometimes you get to understand that some things are not too scary as it sounds, or it seems. And that is helpful also. When it comes to CF, there's few sites I trust, not a lot. This Facebook page and basically Mayo Clinic and the CF Foundation.”}

P22 also noted that she preferred this group because the moderator of this Facebook group \textit{``has strict moderation rules"} where they would delete comments that they thought \textit{``did not serve the community"} or if the comments were \textit{``not mature}.” In some cases, they would also block people from the group if their behavior was found problematic. An example of these guidelines when it comes to COVID is when the moderator selectively removes content that deviates from alignment with cystic fibrosis treatments and viable experiences within the online community. She told us how it reduces her uncertainty to see \textit{“the people writing are parents like Me who want the best for their kids or actual CF patients who are mature and can tell you about their experience in a very good mature way because this group you cannot just say anything. There is like a `manager' in the group and he will block you or He will like, erase your comment. It's just very professional. That's why I like it. And you can see from the comments. When the comments come in, the comments also are very enriching.”}

\subsection{Parental Social Media Burnout and COVID}
While parents in the CF parenting group were acting maturely, partially because of the moderation policies on the Facebook Group, other parents did not have similarly positive experiences in their online interactions especially when it came to COVID. 
In fact, P23 says that \textit{“if [the post] has an exclamation point I step away from it, I need facts not excited feelings.”}  Similarly, P2 gave up using Facebook because of the anti-vaccination comments that were rampant from many of her friends on the site. She emphasized, \textit{“well, if I ever see anything about anti-vaxxing, I'm just like, Ugh. It just-- I used to be very upset about it, very angry. Now I'm just kind of like, Ugh, I just, I don't even want to go there, okay?" So we're all divided on this … So that can be a total turnoff for me now."} P19, a father with one child and another on the way, echoed this perspective, commenting on living with his father who had different opinions. \textit{“My dad is anti-vax..he's definitely on Facebook and [believes in] every conspiracy, [that] more people die from the vaccine than COVID. That it's going to compromise your immune system. That you can get other diseases, like everything out there, which I 100\% don’t agree with, but it's hard to argue [with him].”} P19 does not use any of the popular social media platforms except for LinkedIn. He stopped using Facebook because he did not want to engage with his father on vaccination, masking, and other COVID mitigation issues. He lamented the fact that because of \textit{“our partisanship…everyone's kind of in their bubble right now. And I think technology is partially to blame [with ] algorithms kind of by feeding everyone the information that they want to hear and sort of puts everyone in their echo chamber, I think makes it difficult that we're not able to have those discussions.”} 

However, not all parents thought being in their own bubble was problematic. In fact, P14, a once heavy Instagram user, stopped using the app when comments to her posts became more toxic \textit{“over the course of the pandemic''}, with content appearing \textit{“less genuine and more negative.”} This \textit{“shift”} coupled with the unmoderated negativity made her more inclined to leave the app.  Similarly, P5, who is a self-described \textit{“anti-vaxxer”} and whose young child contracted COVID said she had quit social media because others on her network were insisting that she had to get the vaccine for her child’s benefit. After her child contracted COVID, she argued that, as a young mother caring for her young child, \textit{“ There was no other way for us to proceed, other than how we proceeded. Even if I had known more about COVID and the baby and the likelihood of her getting it... She's got to sleep next to me. We have to nurse throughout the night. I wore a mask as much as I could. So the fact that she had contracted it probably due to nursing, just seemed like, well? [intimating there is nothing she could have done as a mother] So I think that's why I didn't look for information online.”} P2 \textit{“used to love Facebook clips of Stephen Colbert and the Daily Show and all these things where I can get news in a fun way. And those are all gone. Like my algorithm is messed up now. So I probably am checking Facebook less often.”} These clips that P2 used to enjoy browsing through before bed were now replaced by short clips about vaccines, something she did not want to engage in at the end of the long day. 

Other parents found the cavalcade of contradictory information paralyzing. P3 says\textit{“she felt exhausted”} after having to consider so much information from different sources about the vaccine, and how \textit{“the vaccine itself could lead to other side effects in children. And there's not enough studies that, I guess, time has elapsed enough to see the full effects of a vaccine. Things and some things that were true, right, it is true that not enough time has lapsed. And so it kind of leads you down a rabbit hole to think like, "Okay, yeah, what if this? What if that?”} P16, who had two children test positive for COVID twice, expressed a similar sentiment, highlighting the contradictory and contingent information on social media, especially Facebook, during the pandemic. This experience led her to question her relationship with social media and prompted a differentiation between the online world and her physical life, particularly when dealing with COVID. She emphasized the negative emotional impact Facebook had on her family, expressing a desire to break free from the constant anxiety and worry induced by the platform. \textit{“I don't want social media to have this power over me emotionally anymore.”}

\subsection{Parental Concerns with Children's Uncertainty: Online Engagement \& Emotional Response}
We found many parents in our study who mentioned various concerns about their children, who were old enough to have their own types of COVID uncertainty. Children, facing social discomfort during the pandemic, exhibited increased uncertainty, impacting their parents' emotional experiences and responses. The negative impacts of COVID on children's well-being contributed to a challenging dynamic for parents navigating their own uncertainties associated with the ongoing situation. 

Parents mentioned different responses that children had towards COVID. Some children developed a sense of fear through the isolation and the illness symptoms.  
P1 shared with us that her younger child has \textit{“a fear of unmasking, a fear of getting sick. And this was [even before she got diagnosed with COVID]. Like, there were a few times when she started getting sick with various respiratory symptoms, but she would hide it for fear that she would be diagnosed with COVID. And I don't know if there's a stigma that she thinks there's attached to COVID. Even now, there's she has this like fear of being unmasked and getting sick.”} 

P15 tells us how when her daughter contracted COVID, she connected with her school friends using Discord in after-school hours. This discord group acted as a conduit where the middle-school children in her daughter’s class would come together to be socially connected whenever any of them were sick and out of school. She concluded by saying that her daughter’s isolation when she contracted COVID was \textit{“really helped because everybody got on Discord after school… She could still participate in that.”}

However, differences around masking and vaccination policies that we presented in section \ref{sec:networked} also expressed themselves in children’s online interactions. P15’s child \textit{“is one of the few maskers in her class now, and she does it because she doesn't want COVID.”} 

Children on the channel encouraged each other as friends to get tested and wear masks. However, such interactions were not always positive. For example, one of  P15’s school colleagues was coughing at school, so members of the Discord channel were saying that since he \textit{“had symptoms and she's like, "[friend] just test" and he's like, "my parents won't let me." So she's like, "really, just go to the nurse." And was again like, "my parents won't let me." And so they had this impasse.”}

\section{Discussion}
In this section, we first present the theoretical implications of this study, especially as they apply to the Uncertainty in Illness theory (UIT). We then present some design recommendations to provide parents with online communities that better support illness uncertainty management. 

\subsection{Theoretical Implications: Uncertainty in Illness mediated through social media}
\label{sec:discussion_uit}

As with earlier studies that explored modifications to Mishel's original Uncertainty in Illness theory \cite{mishel1988uncertainty}, we propose new factors that can be added to the three segments of antecedents, appraisal, and adaptation in this post-pandemic world. We suggest these as supplementary elements to the theory, offering potential foundations for future work and providing some ways in which it can benefit overall socio-technical research in healthcare. These proposed additions are visually represented in Figure \ref{fig:theory} (with dotted outlines).


\subsubsection{Antecedents: Socio-Technical Factors}
As highlighted in the relevant literature, antecedents are factors influencing the encounter with uncertainty. We contribute to the UIT model by proposing another antecedent component which we refer to as `socio-technical factors'. Online communities significantly contribute to parental perspectives on illness, its severity, and the biological and psychological aspects derived from other parents' shared experiences.

\paragraph{Algorithms}
Many social media platforms are feed-based, made of algorithms that curate posts based on users’ prior interactions \cite{10.1145/2702123.2702556}. Algorithms exhibit varied interactions with individuals, providing some with relevant content they enjoy while exposing others to topics they are attempting to avoid \cite{caseychipaper}. 

Some parents noted that they used to rely on algorithmic recommendations to access information and content that they enjoyed. For example, P2 said that, before COVID, she liked to follow humorous political clips on Facebook as her way to wind down at the end of the day. Similarly, P9 found extensive information on Instagram about unique topics relevant to her such as the "online" lactation consultant she found through Instagram because she engaged more in those posts during her pregnancy before COVID. Algorithmic content curation influences the sequence of posts and comments, which plays a pivotal role in shaping user experiences on social media platforms. These algorithms determine the visibility and order of content, impacting the flow and dynamics of online discussions.

\paragraph{Moderation Governance}

Moderators play a crucial role in upholding and adapting community rules. They actively remove users who disrupt the established ground rules of the community in terms of behaviors and content \cite{shagunmoderation}. There exists an implicit structure where moderators exert significantly more control over the norms and content of the online community than other members \cite{schneider2022tyranny,ammarimoderation}.

Our participants identified moderators as users who are members of the groups, monitor content regularly, and delete anything that does not fit the norms of the group. Removing misinformation becomes an important issue in any medical emergency, especially one as COVID, and it aligns with the understanding that moderation is a required emotional labor that is essential for shaping the identity and values of a community \cite{gillespie2018custodians}. Moderation is more pronounced when creating “non-mainstream forums” for marginalized communities that need stricter security and privacy measures \cite{ammarimoderation, gillespie2018custodians}. Moderators use platform tools to ensure a safe community. They check participants and content eligibility based on profile signals and use features like comment deletion and user expulsion. Moderators play a crucial role in creating safe spaces, controlling discussions, deciding on group participation, and addressing challenges like online harassment and privacy concerns {\cite{ammarimoderation,ammarinetworkempower}}. A good example of this is how P22 respected and trusted moderators on the cystic fibrosis support group to manage a "professional" group, so she does not have to consider the provenance of the data herself. Doing so in fact reflects Richards et al.'s suggestion that medical and healthcare professionals should act as scaffolds for parents \cite{richards_et_al_22}.   

Another type of moderation was identified by parents using the WhatsApp bumper groups, which were created by health professionals for mothers who fit specific criteria. While there were no official moderators in the group, mothers like P8 preferred this group over alternative sources due to its distinctive feature of being established by a verified professional. This instilled trust among the mothers, as they believed the group consisted of individuals who shared content or provided responses based on their experiences with health professionals. Additionally, it was ensured that the group was free from spammers or individuals with different agendas, such as advertisers. Thus, gatekeeping becomes a key factor in deciding whether a parent will be adopting that said channel as an information source \cite{malinen2021boundary,ammarimoderation}. 

\paragraph{Nature of online communities}
Online community platforms often witness the emergence of specialized communities within specific topical areas, characterized by diverse sizes, distinct topical boundaries, and unique rules \cite{10.1145/3512908}. The nature of these communities affects members' behavior. We analyze some examples of distinct online community categories below:

\begin{enumerate}
    \item Algorithmically curated communities: These communities involve information tailored to users' interests, coming from individuals with similar topics of interest \cite{10.1145/3512908, gillespie2014relevance}. These kinds of communities that exist especially on popular social media channels such as TikTok, Instagram, etc. are where the algorithm promotes whatever the user engages in. One example of this as confirmed by P9 was using her Instagram extensively for prenatal chiropractor content before giving birth to her child.
    \item Communities of interest: These communities are where individuals choose to join based on their independent preferences. Communities of interest on social media bring together individuals with shared passions, hobbies, or concerns, fostering connections and exchanges within a specialized topic. These communities provide a platform for like-minded individuals to engage, share information, and build a sense of belonging \cite{shagunmoderation, gruzd2013enabling}. Similarly, parents of children with special needs \cite{ammarinetworkempower}, stay-at-home dads \cite{ammari_schoenebec_fathers_15,ammari_schoenebeck_16}, and LGBTQ+ \cite{blackwell_et_al_16} parents created communities that were focused on their communal challenges. 
    \item Hyper-localized Communities: 
Because of the effective gatekeeping provided by healthcare professionals, WhatsApp "bumper" groups provided a safe space for discussions with other expecting parents facing similar challenges in the same time frame. Mothers like P9 and P6 found each other in this group as they discussed their pregnancy-related issues. The group evolved to become a close community after childbirth for new parenting information, especially that which was necessary while involuntarily transitioning into a pandemic world. This reflects earlier work showing how mothers preferred interacting with other parents on hyper-localized Facebook Groups because of the trust assurances afforded by the link between the virtual space (Facebook Group) and the offline space (local neighborhood) \cite{moser_et_al_17}. 

Similarly, other participants preferred the hyper-localized nature of the community helping to gain information such as daycares, organizing play-dates, etc. in their local neighborhoods. Our findings reflect Felton's study which revealed that the internet is extensively utilized for targeted location-based information seeking in specific populations and how individuals often fuse online searches and information from non-mediated sources \cite{felton2015migrants}.  
    \item Support-specific communities: These communities are where unique groups of people come together to solve the same problem. For example, patients who are veterans contributing to the case of P22's son’s cystic fibrosis. This nature of online communities becomes prominent in parents wanting to become a part of it to reduce, temporarily resolve, or mitigate uncertainty. Similarly, earlier work shows how parents of children with special needs looked for communities that would provide support specific to their needs (e.g., the specific needs of a child on the spectrum in a particular school district) \cite{ammarinetworkempower}. 
 
\end{enumerate}

\subsubsection{Appraisal mediated through social media}
Uncertainty can be appraised as a (a) danger or (b) opportunity and for each of these, an individual can respond with different coping strategies. Below, we describe how coping strategies were mediated by social media use. 

\paragraph{Coping when uncertainty is appraised as ‘Danger’: ‘Online’ Mobilization and Affect-control strategies}
When uncertainty is appraised as Danger (threat), the individual tries to find coping mechanisms such as mobilization strategies or affect-control strategies that reduce or manage their uncertainty. Mishel termed mobilizing strategies as seeking new information or taking direct action \cite{mishel1981measurement, mishel1990reconceptualization, mishel1999uncertainty,mishel1988uncertainty}. Through our findings, we see how parents' mobilization strategies included migrating towards evidence-based information online to reduce their appraised uncertainties about COVID. This parental coping mechanism is taking actions based on information they find in their preferred online community. Parents curated their preferred online communities based on their comfort with other members of the group and the information they were sharing. P15 is a prime example where the parent actively curated her online community on Reddit by engaging with subreddits (r/ZeroCovidCommunity) where individuals shared reliable information and recommendations about continuing COVID mitigations even after most people stopped taking precautions like masking in public.

Appraisal of uncertainty as a danger motivated some parents to find "communities of coping." For example, P22, who already trusted the cystic fibrosis community of which she was a member before COVID, used that space as an uncertainty mitigator. Simpson et. al studied appraisal in individuals with Parkinson's disease in the context of COVID and specified what we see as an echo in our study that if parents have a space already established and the parent is already embedded in the support channel, then they have a better appraisal and coping strategies that promote better adaptation \cite{simpson2022s}. 

Another example is leaning on Hyper-localized Communities for “social distancing”. Parents using WhatsApp groups which were created for exclusive purposes like mothers of babies in the same age range or kids in the same daycare, became “crowd locators'' points to positively navigating through the uncertainty of COVID. This use of local social media outlets also helps in creating a network-level space for mitigating uncertainty. This presents the notion of ‘networked empowerment’ \cite{ammariparentspaper, caseychipaper, ammarinetworkempower}, illustrating how parents connect with one another, access resources, and discover innovative methods to advance health advocacy on both local and national scales.

In addition to mobilization strategies, one might engage in affect-control strategies like practicing emotional disengagement where they intentionally distance themselves from emotional involvement or investment in a situation, relationship, or experience, to manage the emotional distress associated with uncertainty.  In our findings, we see ways in which parents apply this affect-control strategy by reaching out to their “safe spaces” or staying in the hedge of their own trustworthy sources. An example of such is the “Vanderbilt mom” who entrusts only information from her selected medical universities for her children. Another example is P7 who strongly believed that \textit{``people who go to school for certain fields, medical fields, actually know what they're talking about."} She valued their opinions over those of \textit{``a random stranger on the internet."}
 
\paragraph{Filter bubbles: trusted sources of evidence-based and experiential information:}
These parents found or made their filter bubbles which led to confirmation bias. A confirmation bias view on social media induced polarization during COVID \cite{vermadisentangling}. Confirmation bias is characterized by the inclination to interpret, search for, recall, and favor information in a manner that aligns with one's pre-existing beliefs or hypotheses, leading to polarized views on an issue or event \cite{westerwick2017confirmation, modgil2021confirmation}. This cognitive tendency explains why individuals with opposing views on a topic can interpret the same evidence differently. Those affected by confirmation bias tend to assign more significance to evidence supporting their beliefs while downplaying evidence that contradicts them \cite{huang2012understanding,westerwick2017confirmation,zhao2020promoting}.  Our findings suggest that these “generated” filter bubbles can help reduce parents’ COVID uncertainties. It is important to note that when parents appraise it as a danger, the filter bubble operates in one manner, while in the appraisal of opportunity, it operates differently. We discuss this difference in the subsequent sections. 

\paragraph{Coping when uncertainty is appraised as Opportunity: ‘Online’ Buffering Strategies}
When uncertainty is appraised as ‘Opportunity’, the coping strategy that is applied is buffering, or in simpler terms, avoidance. Parents actively utilized hope maintenance methods \cite{mishel1990reconceptualization} to deflect threats to their perception of reality by avoiding new information that could alter their existing perceptions. An example of this was P5, a self-described anti-vaxxer, who despite being bombarded with warnings about COVID transmission to her new-born child if she were not vaccinated, avoided these information sources and accepted what she saw as the unavoidable close contact between her and her child - that they would contract COVID. P2 actively started to avoid Facebook (which she depended on heavily before to relax with funny videos before bed) after experiencing a shift in her browsing feed over the period of COVID because algorithms kept pushing “anti-vaccination jargon” that others in her feed were responding to, even though she herself was not interested in the content. This is an example of how algorithmic mirrors can encapsulate users
within their experience \cite{caseychipaper} - what Gillespie calls recursive loops \cite{gillespie2014relevance} which can lead to symbolic algorithmic annihilation \cite{Andalibi_annihilation_21} of users by reducing the complexity of their needs through grouping them with those closest to them in their network structures and thus introducing algorithmic anxiety into their lives \cite{jahver_anxiety_18}. Harmon-jones \cite{harmon-jones_introduction_2019}[P.19] notes that the ``magnitude of avoidance of new dissonance
is not influenced by the amount of existing dissonance and that spreading of
alternatives occurs before a choice. He proposes changing the definition of dissonance to include the degree to which a behavior will lead to a consequence and the desirability of the consequence.'' The consequences of being infected with COVID, especially as a parent would probably add to the parents' dissonance.

\subsubsection{Adaptation mediated through social media}
Adaptation is the neutral ground \cite{mishel1988uncertainty} that the parent reaches where they learn to live with their uncertainty. Adaptation is when the goal-directed behavior persists, even with the things they wish existed to help them {\cite{mishel1988uncertainty,mishel1990reconceptualization}}.

\paragraph{Positive Adaptation}
In our study, the demonstration of positive adaptation entails finding social support through online channels, thus either maintaining or reducing uncertainty. According to Mishel's Uncertainty in Illness theory (UIT) \cite{mishel1988uncertainty,mishel1981measurement}, a successful adaptation can be that while uncertainty still exists, it is more manageable. In our study, most parents found ways to address their uncertainty; either through actively seeking information from online communities they trusted, or through avoiding information they did not want to access.

\paragraph{Negative Adaptation}
We propose a new component to Mishel’s UIT adaptation which we describe as negative adaptation. Adding socio-technical factors as new antecedent components of uncertainty, and analyzing how they mediate the appraisal of parents’ COVID uncertainties, leads us to believe that some online interactions can in fact lead the parents to a negative adaptation. This “burnout”  is an expression of learned helplessness \cite{seligman1972learned} caused by continuous exposure to high levels of information overload without adequate support. Burnout also removes the locus of control from the parent, as they feel they cannot do things any differently. 
 
We observed that multiple parents “\textit{unplugged}” completely from social media and any kind of online presence. One example of this `unplugging' can be found in P16 when they experienced information overload to a point where they felt disempowered because they could not manage the uncertainty \cite{seligman1972learned}. P16 actively separated her physical life from anything online, relying on her spouse to be the one who uses the internet and returns to her with information that she \textit{``might benefit from"}. 

Experiencing mental exhaustion about the outcomes of COVID was another reason to quit online spaces. Parents chose to disengage from seeking information in online communities as their pre-existing beliefs provided a sense of knowing the expected outcome. Take into account the situation of P19, who lost trust in any form of social media because he attributed the platform to influencing his elderly father, who is at potentially high risk of contracting COVID, to become an anti-vaccine advocate. Parents like P16 shared how if the algorithm consistently directs the parent toward the "same rabbit hole" which the parent has been trying to escape, it results in a negative experience of uncertainty-inducing appraisal by the parent. These parents stopped using social media to access information.

\subsection{Design recommendations}
In this section, we present three design recommendations for online communities that can serve to mitigate parental uncertainty.

\subsubsection{One health portal including "Physician Curated, Community-Backed Checklists"}
The One Health portal, featuring 'Physician Curated, Community-Backed Checklists,' can be designed to provide a credible and community-supported platform for health information that combats all types of pediatric health-related uncertainties. Participants emphasized the need for an all-in-one information hub for parents facing uncertainty, highlighting the collaborative and informative nature of online forums and the advantages of consolidating information, especially pediatric health information, in one place. Parents, particularly P8 and P12, underscored the importance of health checklists for new parents, advocating for fact-based guidelines post-COVID. They pointed out the difficulty of obtaining accurate information without medical guidance and expressed anxiety about misinformation, especially during the early days of the COVID pandemic and their pregnancies. They stressed the necessity for physician verification to ensure reliable and trustworthy health information for new parents, suggesting that a structured and easily accessible checklist could be beneficial in navigating uncertainties and avoiding potential misinformation about COVID, parenting, and pregnancy.

The proposed portal will offer numerous benefits, emphasizing the potential impact of the portal on both parents and doctors. It will allow parents to maintain comprehensive health records for their children, ensuring easy access to all medical information. For physicians, it provides a platform to create support groups, similar to the WhatsApp bumper groups mentioned in our findings. The portal also includes pediatric care checklists, fosters a sense of community, and enables parents to share experiences, seek advice, and gain support for their uncertainties in real-time.

Health professionals can upload a wide range of resources specifically designed for children with special needs and create a community with veterans of specific conditions. Much like P22, who found specific resources for her son with CF from her existing online CF community, that ensured access to reliable and relevant information even during the COVID pandemic. Health professionals can also include care plans and educational materials. The portal could be integrated with popular electronic health record viewers like MyChart, streamlining the tracking of health metrics and care plans and enhancing the overall management of a child's health.

Moreover, the portal could feature contributions from a global community of medical professionals specializing in various medical contexts, ensuring that the latest research and best practices are shared widely, benefiting patients worldwide. Functioning like a Reddit for medical professionals and parents, the portal would be viewable by patients and moderated by doctors or physician assistants, promoting and facilitating the exchange of knowledge and ensuring that information is accurate and up-to-date.

The portal is designed to function as a social, community-centered, healthcare professional-moderated online space. The physician-led and community-supported checklists will be backed by the evidence-based information that parents seek. Based on our findings, physicians and other healthcare professionals can assess the needs of the local community and, in some cases, take it upon themselves to moderate and contribute to the online communities in this portal.

This portal can also have hyperlocalized affordances, enabling the formation of groups that help parents seek support within their area, similar to the WhatsApp bumper groups that participants in our study utilized more than other online spaces. We envision creating this portal to provide the social network affordances of social media platforms while incorporating many of the essential features of current medical portals. These features include comprehensive health record management, physician-verified information, and specialized resources for various health conditions. By merging the community engagement aspects of social media with the reliability and functionality of medical portals, this platform aims to support parents and healthcare professionals in a comprehensive and trustworthy manner.

\subsubsection{Online Health Professional Credibility Identifier}

Parents turn to online communities for assistance and support, scrutinizing information credibility at each step of their quest for understanding. Their preferences for specific communities, as discussed in earlier sections, hinge on the source reliability, testimonies, and observed outcomes in other parents' lives. These are queries that health professionals can address with authority. Therefore, establishing online groups and communities led by medical professionals (as the "bumper" WhatsApp groups were created) or with their active involvement (as with the CF group described by P22), can alleviate uncertainty and foster improved health efficacy practices. We propose incorporating verified accounts of health professionals in parenting support groups. Following models of verification currently in use on platforms like Facebook, this study suggests the use of a verified and unique badge or identifier that clearly indicates a person's status as a medical professional, including their relevant skills and current employer. The aim is to enhance credibility and facilitate easy identification of trusted health professionals within the community.

In this scenario, medical professionals' involvement in online communities may lead to increased demands on their time and resources. They might need to allocate significant time to monitor, contribute, and provide guidance within these groups. Additionally, there is a risk of misinterpretations. To mitigate this, implementing disclaimers that clearly define the scope of advice provided online can be helpful. It is also beneficial for medical professionals to refer participants to appropriate in-person medical services when necessary. This design component calls for future work to develop efficient methods for balancing online engagement with professional responsibilities, ensuring accurate information dissemination while maintaining the integrity of in-person medical care.

\subsubsection{Personal moderation tools and sentiment `gauges’ for parents in online communities}
Earlier studies review simulated personal moderation interfaces which allow users to customize their moderation beyond what is generally provided at the platform level \cite{shaguncontentmoderation,shagunmoderation,gonccalves2023common}. We propose this for the online communities we identified in our findings. Parents stepped away from posts with “exclamation marks”, “excited feelings”, and “posts with negative hashtags”. Personal content moderation can be effective in mitigating uncertainty within online spaces, specifically, at an intersection of sentiment and areas of discourse \cite{gonccalves2023common}. For example, parents might want to avoid negative posts about vaccines in their feeds. 

Sentiment analysis of posts with a display showing positive or negative scores can act as a ‘sentiment gauge’ which shows parents a percentage score of how positive or negative the post, or comment is. Parents can then customize their own preferred score so that they have the option to select what they view, and what may not be suitable for them to see will be filtered out automatically. Analyzing sentiment-revealed follower negativity will offer insights beyond standard metrics on social media \cite{alsayat2022improving,poecze2018social}.

\section{Limitations and future work}
The study comes with certain limitations that can serve as a foundation for future research investigations. First, marginalized cohorts, BIPOC parents, parents characterized by low socio-economic status (SES), and individuals with restricted educational backgrounds represent potential participants for further exploration through the application of UIT with socio-technical factors. Other works often address the challenge of recruiting individuals from low socioeconomic status (SES) backgrounds for studies \cite{wambua2022lessons,sun2022battle}.
The coping strategies we discussed in our results may differ among such individuals, based on the common assertion that misinformation is more prevalent in populations with lower educational levels \cite{Crouse, Tang}.
We acknowledge this as a limitation in our study and intend to collaborate with non-governmental organizations (NGOs), agencies \cite{yancey2006effective}, and even hospitals \cite{wambua2022lessons} that can assist in recruiting such participants for future studies.

Additionally, we recruited 18 participants who identified as females and 5 participants who identified as males. The number of fathers was a limitation in our study. Recruitment of fathers is a general challenge that has been addressed in various studies. Various studies discuss how fathers are constrained by privacy concerns and concerns about being judged when sharing content online about their children \cite{ammari_schoenebec_fathers_15, matthews2022fathers,mitchell2007conducting,yaremych2023recruiting}. Moving forward, we aim to over-sample male participants. To achieve this goal, we may explore avenues such as parental networks or electronic health records with the assistance of liaisons such as practitioners.

Further, our recruitment process primarily attracted birth parents or biological parents. Despite our efforts to include a diverse range of parent groups (adoptive parents, foster parents, grandparents who have been caregivers to orphans, etc.) through recruitment messages on the flyers and social media, our study participants were predominantly birth parents. According to US Adoption Statistics, adoptive parents make up only 2\%—4\% of the United States population. Participation from these groups was minimal due to their low base rate in the population. 

In addition to these, future research endeavors can construct a psychological metric based on socio-technical factors to enhance the quantification of the discoveries made in this study. Utilizing our study as a theoretical framework for future studies would be valuable. 

Finally, we only examined the impact of uncertainty on parents. Healthcare professionals and practitioners experience their own set of uncertainties about novel viruses and diseases like COVID. Future studies should address how online community use intervenes with the uncertainty of healthcare professionals.

\section{Conclusion}
In this qualitative study, we interviewed twenty-three parents of children who contracted COVID to better understand how their use of social media facilitated their uncertainty management. We found that technological factors have significant effects on whether parents were able to access the resources (support and information) they needed. We propose new components to the original theories on uncertainty in illness and its re-adaptions and provide several design recommendations to support parents in their quest to manage health risks and better navigate pediatric uncertainties.

\section{Acknowledgments}

\bibliographystyle{ACM-Reference-Format}
\bibliography{sample-manuscript}

\section{Appendix}

\subsection{Interview Questions} \label{sec:appendix_interviews}

\section*{Section 1 - About the caregiver, family, and home dynamics}

\begin{enumerate}
    \item Tell me about your family. How many children do you have?
    \item Who lives with you in your house?
    \item Who in your home is currently employed? What do they do?
    \item Describe a typical day from morning to evening in your home (Probe: how do your children get to school or other activities?)
\end{enumerate}

\section*{Section 2 – Children’s Health, existing conditions and COVID Past}

\begin{enumerate}
    \setcounter{enumi}{5}
    \item How old is your child? When did you first suspect your child had COVID?
    \begin{enumerate}
        \item Were they infected multiple times?
        \item Did your kid have any Post-COVID conditions?
        \item Probe: Which vaccines? / For what illnesses?
        \begin{enumerate}
            \item Gauge/ don’t probe: How is the child doing now?
        \end{enumerate}
    \end{enumerate}
    
    \item If your child has a health concern, who usually manages it?
    \begin{enumerate}
        \item Can you talk briefly about how you and your partner share a child’s healthcare responsibilities?
    \end{enumerate}
    
    \item Have you had any differences with [your partner] about treatments of your child?
    
    \item What kind of changes did you see in your daily activities when your child/children had COVID?
    \begin{enumerate}
        \item Also, any changes, Post-COVID?
        \item When did you /did you ever start suspecting your child has long-term effects of COVID-19?
        \item What/Which sources of information did you find online? Who did you discuss it with?
        \item Do you have any support groups online?
        \begin{enumerate}
            \item Probe about where the group is, how they found it, and how often they use it.
            \item Probe What do you (does your partner) do to look for information about the condition, treatment plans, etc.?
        \end{enumerate}
    \end{enumerate}
 
\end{enumerate}

\section*{Section 3 – Social Media Engagements}

\begin{enumerate}
    \setcounter{enumi}{4}
    \item What social networking sites are you/your partner on?
    \begin{enumerate}
        \item Do you post statuses or photos about your children on social networking sites?
        \begin{enumerate}
            \item If yes, what do they say? /what’s the most recent one?
        \end{enumerate}
        
        \item Can you share a little bit about how social media has factored into you being a parent?
        
        \item How has that changed at the time your child contracted COVID/post-COVID?
    \end{enumerate}
    
    \item Do you read other parents’ statuses or look at their photos of children? Do you engage with their posts in any way? (like, comment, share, etc.)
    \begin{enumerate}
        \item How has that changed at the time your child contracted COVID/post-COVID?
    \end{enumerate}
    
    \item Can you share some ways in which social media might have intersected with your thoughts or ideas about the following:
    \begin{enumerate}
        \item Treatment plans – Did you research them online for your child?
        \item Practitioner’s advice – Did you look up any information that the practitioner might have mentioned about your child’s treatment or condition?
    \end{enumerate}
\end{enumerate}

\section*{Section 4 – Trust in Social Media Engagements}

\begin{enumerate}
    \setcounter{enumi}{8}
    \item How do you decide whether to trust the parenting information you see online?
    \item Can you remember a time when something from a social media post stood out to you and made you change the way you care for your child?
    \begin{enumerate}
        \item Did you change the way you interacted with others who cared for your child?
        \item Was there ever something you learned online about COVID that made you do something or not do something for your child? (try to assess their trust)
    \end{enumerate}
\end{enumerate}

\section*{Section 5 – Parental Burnout}

\begin{enumerate}
    \setcounter{enumi}{10}
    \item What has made you most emotional about your child’s care?
    \item During your child’s care and treatment, how would you describe the change in frequency and content of your social media interactions?
    \item How have the responses to your social media interactions made you ever feel?
    \begin{enumerate}
        \item How have these interactions made you more anxious?
        \item Did you change your mind regarding your child’s care because of anxiety (caused by something you saw online)?
    \end{enumerate}
\end{enumerate}

\section*{Section 6 – About Practitioners}

\begin{enumerate}
    \setcounter{enumi}{13}
    \item Can you think of a time when something you read online gave you concerns that you might have shared with your practitioner?
    
    \begin{enumerate}
        \item Can you give me some examples?
        \item What was the reaction from the provider?
        \item Can you think of a time where you experienced discomfort about something when talking to your practitioners about your child’s health/treatment/care?
        \begin{enumerate}
            \item If so, do you discuss these issues with other parents/caregivers online?
        \end{enumerate}
    \end{enumerate}
    
    \item Have you ever looked up what your child’s medical practitioner shared with you about the child’s treatment plan, medications etc.?
    \begin{enumerate}
        \item Which sites/posts/influencers did you look up?
        \item Have you discussed the medical plans with other parents online?
        \item Who among your social network do you discuss this with most?
        \item Did that influence your view of your practitioner?
    \end{enumerate}
    
    \item How have your online discussions changed the way you interact with other adults who care for your child? (if they ask who? – teachers, coaches, librarians, school bus drivers, daycare, other parents, family members?)
    
    \item In what ways did you explain to your child what was happening to them?
    \begin{enumerate}
        \item Probe about Masking
        \item Probe about Child’s perception of vaccines
        \item In what ways did you explain [insert topic] to your kids?
    \end{enumerate}
\end{enumerate}

\subsection{Code Book} \label{sec:appendix_codebook}

\begin{itemize}
    \item Home Life (descriptions of the daily routines of the family and the dynamics of family members interactions)
    \begin{itemize}
        \item Types of Household: (discussion about changes for the children when they are living with either parent who are separated, or living in a home with grandparents) 
        \item Main Caregiver (discussions about the main decision-maker in health decisions for the child. In other words, defining who the primary caregiver is in the family unit and how that affects the dynamics of the family)
        \item Schooling (discussions included child’s schooling habits, commute to school, school mandate changes during lockdown, daycare and other affected activities)
        \item Daily Routines (discussions about activities in the daily routine of caregiving, other illnesses or past medical histories)

    \end{itemize}
    \item COVID infection (discussion, diagnosis, and navigating lockdown, isolation, other household member’s safety)
    \begin{itemize}
        \item Household infection events (description of first infection in members of household, questions about when child was first infected, how was the illness explained to the children)
Multiple infection events were discussed with the interviewee to determine how they changed the family dynamics.

For example, if a partner had been infected before the child, that gives them some personal experience about COVID. Another example is an infection of a child in their social networks.

        \item Infection severity
        \begin{itemize}
            \item Child (discussion of how severe the infection was and whether there were any long-term effects for the child)
            \item Other household members (discussion of how severe the infection was within the household, ways they tackled multiple members testing positive at the same time)
        \end{itemize}
    \end{itemize}
\end{itemize}

\begin{itemize}
    \item Information sources (source from which caregivers primarily found their information)
    \begin{itemize}
        \item Online 
        \begin{itemize}
            \item Whatsapp groups created by healthcare providers (discussion about moderation, being part of a group created by a Healthcare Provider and thus being deemed safer than other online groups such as on Reddit or Facebook)
            \item Reddit Groups (discussion about online channels and other parents sharing “experience-based” information)
            \item Online Parenting influencers (famous personalities that suggest various parenting styles and skills, prominent especially with new parents)
            \item Facebook Groups (discussion about the dynamics of Facebook groups especially interactions with other parents. Parents discussed the types of posts they shared, and how their interactions changed during COVID lockdowns. Discussion also included engagement in posts by other parents, sharing, resharing of “experience-based” information)
            \begin{itemize}
                \item Moderated Facebook groups (discussion about posts that were offensive, not relevant to the group etc., being removed or controlled by moderators)
                \item Algorithmically controlled feed on Facebook (discussion about posts that appeared on the feed not by choice but by algorithms)
                \item Existing online circles (existing groups and support circles parents were a part of that became scaffoldings for accurate healthcare information during the pandemic)
            \end{itemize}
        \end{itemize}
        \item Healthcare Providers (discussion on receiving information during and after patient-provider interactions)
        \item Google Scholar (discussion about using research articles over online sources)
        \item Social Circle (discussion about discussing healthcare with family members and people outside the immediate family connection) 
        \begin{itemize}
            \item Family (discussion about discussing healthcare choice, vaccination ideologies with family. Discussions also pivoted towards avoiding large family gatherings, hyperlocal events at daycares, libraries etc.)
            \item Extended Family that work in Healthcare (discussion about health information shared by those who worked in healthcare within the extended family, reaching out to them for healthcare question rather than reaching out to healthcare workers or online channels) 
        \end{itemize}
        \item News / Blogs / Media (discussion about finding information from live broadcasts, newspapers, etc)

    \end{itemize}
    \item Trust (discussion about trust levels of caregivers for assessing the accuracy of healthcare information)
    \begin{itemize}
        \item In their social network (discussion about trust in information about illness severity, treatments, precautions etc. happening with family, neighbors, daycares etc.)
        \item In Healthcare Providers (discussion about trust in information provided by the healthcare providers about vaccinations, preventive measures etc.)
        \item In different online health information sources( For example: WebMD, Mayo Clinic etc) 
Research articles via Google Scholar (discussions emerged about using google scholar or following citations in other articles that led to scholarly published work) 

    \end{itemize}
    \item Changes (discussion about emotional and/or behavioral shifts before, throughout, and after COVID infection)
    \begin{itemize}
        \item In trust (discussion about what they heard from healthcare providers, what they saw online and what they heard from their personal networks.  For example: as reflected in findings, some parents turned to online sources and assessed authenticity of that information based on “experience of other parents” because their own providers or pediatrician would “tell something different every time” )
        \item In behavior (discussion about changes or certain measures they might have taken to assess or cope with uncertainty, anxiety etc. For example: turning to whatsapp group more for information than calling the healthcare provider)
        \item In habits or attitudes (discussion about measures they took to deal with information that led to uncertainty and other emotions. For example: one way to avoid COVID information is not using Facebook anymore; another example exemplified the behavior of actively searching for information by developing the habit of reading research articles by specific organizations like universities and/or hospitals)

    \end{itemize}
    \item Vaccinations (discussion about attitudes towards vaccinations before COVID and ideological stances about the vaccine in the case of COVID) 
    \begin{itemize}
        \item Family member vaccination Status (discussion about who in the family is vaccinated and who had anti-vaccination views of the vaccine before/after COVID. For example, one family could have a parent who was pro-vaccine but a grandfather who was an anti-vaxxer)
        \item Child’s vaccination status (discussion about side-effects, if any)
        \item Attitudes towards vaccination (discussion about choice of taking the vaccination, probed deeper to find motivation e.g: if it was for their safety, or fear of infecting others) 
        \item Worries about vaccination (discussion about side-effects in adults, information they might have found through the various information sources listed above)

    \end{itemize}
    \item Masking (discussion about masking, different rules around masking e.g: Children under 2 did not need masks on an airplane) 
    \begin{itemize}
        \item Masking in schools
Children’s uncertainty about masking (discussion about how children age 3 and under perceived masking, parents anxieties about children not being able to recognize a masked parent)

        \item Masking in public
Masking techniques from online channels (discussion where caregivers found different ways to promote masking in children like buying masks showing their favorite cartoon characters)

    \end{itemize}
\end{itemize}

\begin{itemize}
    \item Children’s uncertainty and Illness (discussion about how parents perceived their children’s assessment of uncertainty as young patients) 
    \begin{itemize}
        \item Child’s coping mechanisms both offline and online (discussion about children who were old enough to have a social media presence and found ways of social bonding online. Specifically, the focus is on their discussion of COVID-related issues)
        \item Difference between illness severity in siblings in the family or other children in the wider social network (any illness comparisons mentioned by the parent about their kids, among siblings, within classes at school or daycare)
    \end{itemize}
\end{itemize}

\end{document}